\documentclass[sigconf]{acmart}

\AtBeginDocument{%
  \providecommand\BibTeX{{%
    \normalfont B\kern-0.5em{\scshape i\kern-0.25em b}\kern-0.8em\TeX}}}

\acmConference[PACT '22]{PACT '22:  International Conference on
Parallel Architectures and Compilation Techniques (PACT)}{October 10--12, 2012}{Chicago,IL}
\acmBooktitle{PACT '22: International Conference on
Parallel Architectures and Compilation Techniques (PACT), October 10--12, 2022, Chicago, IL}
\acmPrice{15.00}
\acmISBN{978-1-4503-XXXX-X/18/06}

\usepackage{algorithm,algpseudocode}
\usepackage{amsmath,amsthm,enumitem}

\newcommand\draftmod[1]{\textcolor{black}{#1}}

\newcommand\draftmodsec[1]{\textcolor{black}{#1}}

\newcommand\draftmodfinal[1]{\textcolor{black}{#1}}

\newcommand\draftmodfinalremove[1]{}

\newcommand\Xlabel[0]{VoxelCache}
\newcommand\Xinstructionprefix[0]{voxcache}

\newcommand*\circled[1]{\tikz[baseline=(char.base)]{
            \node[shape=circle,fill,inner sep=1pt] (char) {\textcolor{white}{#1}};}}

\usepackage{subcaption}
\usepackage{tikz}
\usepackage{xcolor}

\renewcommand\footnotetextcopyrightpermission[1]{} 
\begin{document}

\title{\draftmodsec{\Xlabel: Accelerating Online Mapping\\ in Robotics and 3D Reconstruction Tasks } }

\author{Sankeerth Durvasula}
\affiliation{%
 \institution{\emph{University of Toronto}}
\country{Canada}
}

\author{Raymond Kiguru}
\affiliation{%
 \institution{\emph{University of Toronto}}
\country{Canada}
}

\author{Samarth Mathur}
\affiliation{%
  \institution{\emph{PES Institute of Technology}}
\country{India}
}

\author{Jenny Xu}
\affiliation{%
 \institution{\emph{University of Toronto}}
\country{Canada}
}

\author{Jimmy Lin}
\affiliation{%
 \institution{\emph{University of Toronto}}
\country{Canada}
}

\author{Nandita Vijaykumar}
\affiliation{%
 \institution{\emph{University of Toronto}}
\country{Canada}
}

\renewcommand{\shortauthors}{Durvasula, et al.}

\begin{abstract}
Real-time 3D mapping is a critical component in many important applications today including robotics, AR/VR, and 3D visualization. 3D mapping involves continuously fusing depth maps obtained from depth sensors in phones, robots, and autonomous vehicles into a single 3D representative model of the scene. Many important applications, e.g., global path planning and trajectory generation in micro aerial vehicles, require the construction of large maps at high resolutions. 
In this work, we identify mapping, i.e., construction and updates of 3D maps to be a critical bottleneck in these applications. The memory required and access times of these maps limit the size of the environment and the resolution with which the environment can be feasibly mapped, especially in resource constrained environments such as autonomous robot platforms and portable devices. To address this challenge, we propose \Xlabel{}: a hardware-software technique to accelerate \draftmodsec{map data access times} in 3D mapping applications.
We observe that mapping applications typically access voxels in the map that are spatially co-located to each other. We leverage this temporal locality in voxel accesses to cache indices to blocks of voxels to enable quick lookup and avoid expensive \draftmodsec{access times}. We evaluate \Xlabel{} on popularly used mapping and reconstruction applications on both GPUs and CPUs. We demonstrate an average speedup of 1.47X (up to 1.66X) and 1.79X (up to 1.91X) on CPUs and GPUs respectively.
\end{abstract}



\maketitle

\section{Introduction}
3D mapping is an essential component in many important applications today, such as in autonomous robotics, AR/VR applications, and 3D reconstruction. Mapping involves the real-time construction of a representation (map) of the environment. This is typically done by continually processing and integrating distance information to objects in the scene that is obtained from sensors such as depth cameras and LiDAR. Mapping is an integral part of tasks such as SLAM (Simultaneous Localization and Mapping), which involves determining the location/pose of a mobile robot within the constructed map of the environment, and 3D reconstruction, which is used to construct and render 3D scenes/objects from 2D images. These tasks are deployed in a number of important applications. For example, augmented reality libraries for mobile phones like Google's AR-Core~\cite{arcore} and Apple's ARkit~\cite{arkit} use SLAM for motion and environment estimation. 3D Reconstruction is used for a number of visualization tasks, like 3D renderings of real estate floorplans for apartment rentals, and free viewpoint television, which enables interactively viewing large city models from different any viewpoints used for street view.

One common approach to represent the environment as a map is to use a voxel grid, in which the 3D space is discretized into a grid and each element of the grid (called a voxel) stores local information that characterizes that point in space. Typically, the distance to the closest occupied point in space is saved in each voxel, referred to as the Signed Distance Field (SDF)~\cite{sdf}. Mapping continually updates the information in each voxel by integrating the depth input information obtained from sensors with repeated calls to a map update routine. It is often necessary to have both fast map updates that can process high input frame rates and generating maps at high resolution (more voxels to represent a given region). For example, in robotics perception, faster updates allow for a faster control loop feedback which enables robust and agile maneuvering. In online scene reconstruction, a fast and high resolution map allows a more detailed render to be generated at a higher frame rate. In autonomous navigation, a higher resolution map would result in the application inferring finer details of its surroundings, enabling more optimized trajectories during path planning \cite{robust_replanning,realtime_replanning_bsplines,raptor, helenol_clutteredenv,helenol_traj}.

However, enabling fast map updates at very high resolutions is challenging, especially for mapping large environments. We find that a single map update takes anywhere from around $50ms$ to $500ms$, depending on the required resolution of the map. This is due to both lengthy access latencies and a large number of accesses to voxel data per map update. Furthermore, higher map resolutions significantly increase the number of voxels required to be updated. For example, to have a $2X$ resolution requires each map update to access and update $8X$ the number of voxels.

\draftmodsec{Thus, mapping large environments necessitates \emph{fast} access to information stored in voxels for vast stretches of the environment in memory in a \emph{scalable} manner}. Voxel hashing~\cite{voxel_hashing} \draftmodsec{is a commonly used technique} that offers a memory efficient representation while having reasonable access latencies to voxel data, making it ideal for a number of high speed mapping tasks to dynamically represent large areas of the environment~\cite{voxblox,voxgraph,fiesta,openchisel,bundlefusion,voxel_hashing,voxhash_hierarchical,mesh_hashing,refusion,infinitam,flashfusion,voxhash_fastmeshing,gpu_reconstr}. It uses a hash table to map coordinates in space to an $n\times n\times n$ block of voxels (voxel block). However, we find that the percentage of time that is spent in accessing voxel SDF information from the voxel hash is up to $60\%$ of the overall map update time, leading to high latency map updates. \draftmodsec{Another commonly used technique is to use octrees~\cite{octomap}. State-of-the-art octree implementations store blocks of voxels as leaves of a tree data structure, but still incur similar voxel data access latencies as that of voxel hashing~\cite{efficient_largescale_reconstruction}.}

Our goal is to accelerate this voxel data retrieval time during map updates in order to have the low latency map updates at high-resolution that are crucial for many important applications. We observe that while mapping requires a large number of updates/accesses (in the order of millions) to voxels, many of which are repeated accesses to the same voxels.

We introduce \Xlabel{}, a hardware-software mechanism to enable fast map updates by ensuring low latency accesses to voxel data. The key idea behind \Xlabel{} is to leverage temporal locality in accesses to the voxels by caching the pointer to the voxel corresponding to each 3D coordinate. \Xlabel{} enables fast lookups to the voxel blocks by using the on-chip caches to save the voxel block pointers for recently accessed voxels. This avoids the need to perform expensive hash table computations for each lookup \draftmodsec{ in the case of voxel hashing or expensive tree traversal in octrees.} 

We demonstrate that our approach achieves speedups of $1.47X$ (upto $1.66X$) and $1.79X$ (upto $1.91X$) on average in mapping update for various commonly-used SLAM and 3D reconstruction frameworks on CPUs and GPUs respectively. The contributions of this work are as follows:
\begin{itemize}
\item We perform a detailed characterization of various 3D mapping and reconstruction tasks, and we observe that voxel hash resolution forms a significant performance bottleneck for mapping. 

\item We introduce \Xlabel{}, a hardware-software mechanism to accelerate mapping updates at higher resolutions by enabling faster accesses to voxel data.

\item We evaluate \Xlabel{} for various SLAM, online 3D reconstruction frameworks that construct an SDF map of its environment on both CPUs and GPUs. We find that \Xlabel{} is able to provide significant speedups in map update latencies for these workloads.
\end{itemize}

\section{Background and Motivation}
\subsection{Representation of Voxel Maps in Memory}
A straightforward approach to store voxel data is to use a 3D array indexed by the spatial coordinate of the voxel~\cite{kinectfusion,kintinuous}. However, this approach is not scalable and becomes impractical for large maps or at high resolution as it requires a significant amount of memory for storage. For example, a 3D array to represent occupancy information for a relatively small $10m \times 10m \times 10m$ region with voxels of resolution $2cm$ would require $500MB$. 

\subsubsection{\draftmodsec{Voxel Data Structures}}
\textbf{Voxel hashing}~\cite{voxel_hashing} provides an efficient way to store maps of large areas in a scalable manner while incurring low access/insertion latencies. Compared to a 3D array representation of voxels which densely represents both empty and occupied spaces, voxel hashing only allocates voxels for areas that are occupied. Unallocated voxels are considered to be empty making it highly scalable with respect to memory. Voxel hashing thus involves saving allocated blocks of voxels (of size $n\times n\times n$) that is indexed the coordinates of voxel block using a hash table. Mapping a large scene involves a number of allocations of blocks of voxels. A large block size would decrease the number of hash table entries needed, but would greatly increase the memory footprint occupied by the block. Since the target platforms for a number of SLAM/mapping and reconstruction frameworks are embedded systems which have constraints on memory and compute power, a large block size is generally not preferred. A block of dimensions $8\times8\times8$ is usually chosen.

\draftmodsec{\textbf{Octrees}~\cite{octomap} are another popular approach for storing voxel information of large regions of the environment in memory. Octrees use a tree data structure to store voxel data in space. Each node in the tree corresponds to a cubic volume in 3D space. A node may contain $8$ children, which represent 8 equal octants of the cube, or it may be a leaf node, which corresponds to an individual voxel. An access to a voxel involves traversing through the tree data structure starting from the root node to the leaf node. Since each access to a voxel requires a tree traversal, octrees incur a high access latency, making online mapping a lot more time-consuming compared to implementations using voxel hashing. However, some recent works on online mapping with octrees propose specialized implementations that achieve similar performance as implementations with voxel hashing~\cite{efficient_largescale_reconstruction, octtree_fusion_scalable}}.

\begin{figure}
    \centering
    \includegraphics[width=0.8\linewidth]{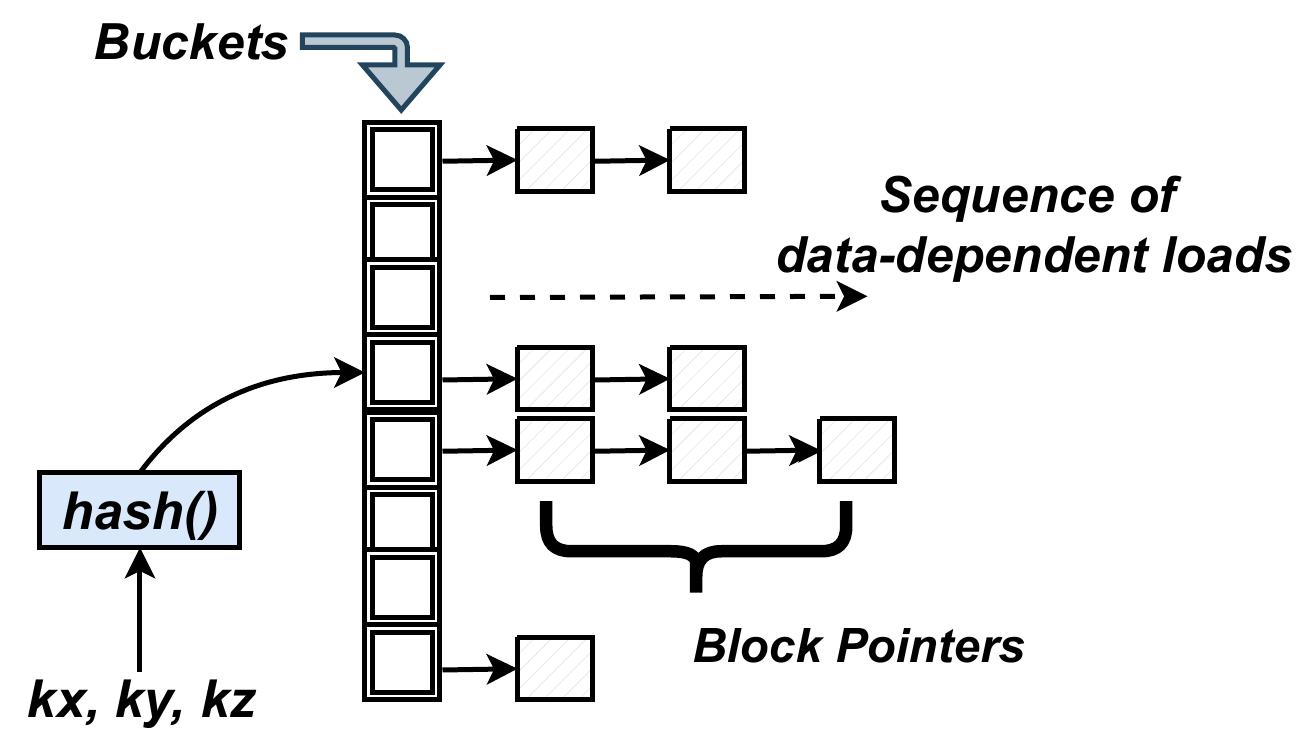}
    \caption{Retrieving block pointer with a hash table lookup}
    \label{fig:HT_access}
 \vspace{-15pt}
    
\end{figure}

\subsubsection{Signed Distance Fields and Map Updates} 
\label{section: map_update_descr}
Signed distance field (SDF) information stored in voxels have previously been used to represent 3D volumes in graphics applications, but are increasingly deployed in computer vision and simultaneous localization and mapping (SLAM) tasks due to the flexible representation and accurate 3D reconstructions it is able to provide~\cite{sdf, kinectfusion}. A signed distance is computed for each voxel in the map, and represents the estimated distance to the closest occupied surface in space. While signed distance fields provide high quality environment representation, they have large memory requirements. A voxel grid map to store the signed distance is the most commonly used approach in mapping literature~\cite{kinectfusion,refusion,voxgraph,kintinuous,voxblox,infinitam,cblox, octtree_fusion,mesh_hashing,openchisel,fiesta}. In this work, we mainly consider frameworks whose map construction maintains and updates signed distance information stored in voxel grids.

The map update routine receives depth measurement information of different points in space from either depth cameras / LiDAR or stereo vision cameras, and a pose estimate of the sensor from odometry measurements or external pose measurements as the input. \draftmodfinalremove{A standard map update routine involves raycasting, where rays are cast from the sensor origin to each observed point. Each voxel intersected along this ray updates its value to the signed distance between the voxel center and the depth of the observed point, up to a truncation radius, resulting in the estimation of the Truncated Signed Distance Field (TSDF). Starting from the center of the occupied voxel along each ray, the signed distance is computed for all surrounding voxels up to a clear radius as the minimum distance to an occupied voxel, thus estimating the euclidean signed distance field (ESDF).} \draftmodfinal{Real time computation of true SDF from depth sensor measurements over a large volume is computationally expensive~\cite{kinectfusion}. One common approach to obtain an approximate SDF is to compute the truncated signed distance field (TSDF) in which the point cloud captured by the sensor is first transformed from the camera's coordinates to the global coordinates. Multiple rays are cast from the camera center to each point in the point cloud. The distance from each point to surrounding voxel centers along the ray up to a truncation distance is calculated and assigned to these voxels. Another approach to estimate an approximate SDF is to compute the Euclidean Signed Distance Field (ESDF) in which the distance from each occupied voxel center (closest voxel to each point in the point cloud captured by the sensor) to all neighboring voxels up to a clear radius is computed. These distances computed are used to update the signed distance of corresponding voxels. }

\subsection{Characterizing Mapping Overheads}
\begin{figure}[htb!]
\centering
\includegraphics[width=1.05\linewidth,trim={0 0.7cm 0 0.5cm},clip]{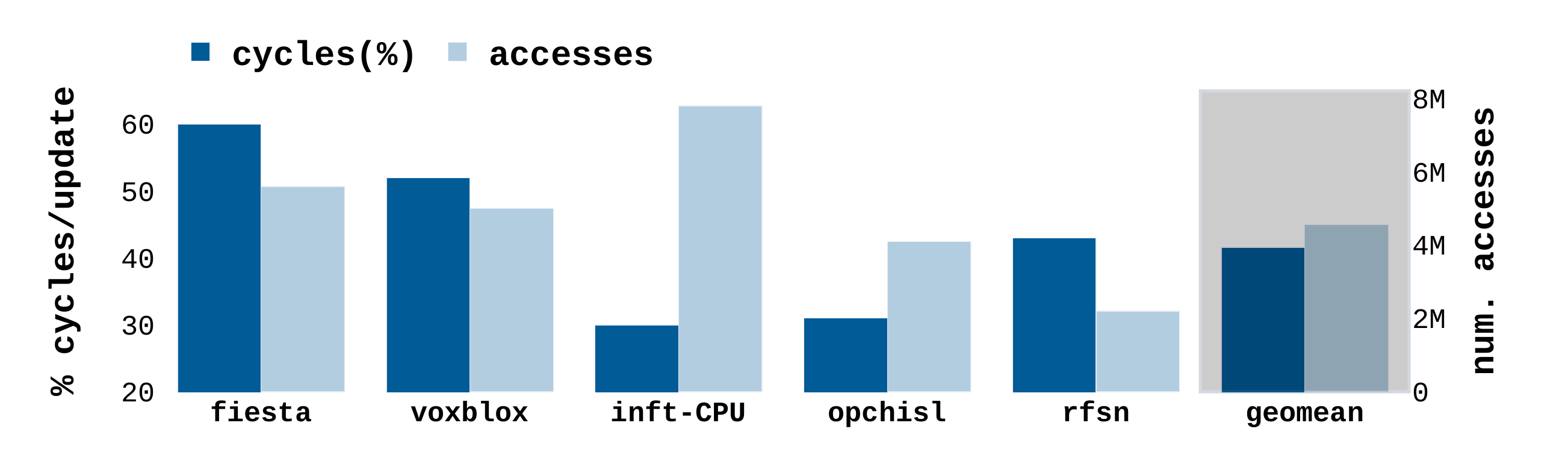}
\caption{Analysis of the number of voxel block accesses and fraction of overall runtime attributed to voxel block accesses}
\label{fig: accesses and cycles}
\end{figure}

Figure \ref{fig:HT_access} illustrates how a voxel block pointer is retrieved from its coordinates by computing the hash function and iterating through a data-dependent sequence of load and key comparison operations. This is a high latency operation because of data-dependent memory accesses and the high branch mispredict rates in the control flow. A large number of these operations are performed at each map update leading to a significantly higher map update time. These operations form a significant portion of the overall map update time. An analysis of the contribution of this latency as a percentage of runtime and absolute number of accesses to the voxel hash table is shown in Figure \ref{fig: accesses and cycles}. \draftmod{Note that in the case of \texttt{infi-CPU}, despite the relatively high number of accesses we observe a relatively smaller percentage of cycles contributing to the map update time. This is because \texttt{infi-CPU} uses a memoization scheme in software which caches the recently accessed block pointer at each hash table bucket. While reducing the number of hash table accesses, it incurs additional overhead for memoization at every access.} 

\begin{figure}[htb!]
    \centering
    \includegraphics[width=1.1\linewidth,trim={0.0cm 0.cm 0.0cm 0cm},clip]{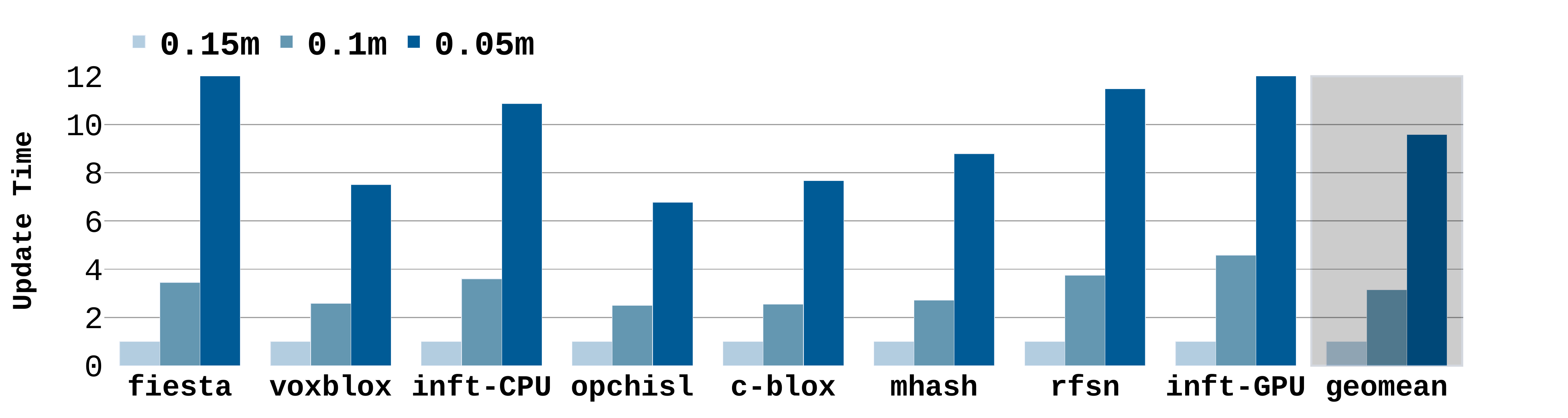}
    \caption{Normalized map update time with varying voxel grid size}
    \label{fig: resolution_update_times}
\end{figure}

At higher resolutions, the number of accesses made at every update increases greatly, leading to a significant increase in map update time. Figure \ref{fig: resolution_update_times} shows the impact of resolution on map update time. On average, the map update time increases by $9X$ between grid sizes $0.15m$ and $0.05m$.

\begin{figure}[htb!]

    \centering
    \includegraphics[width=.9\linewidth,trim={0 0.3cm 0 0.5cm},clip]{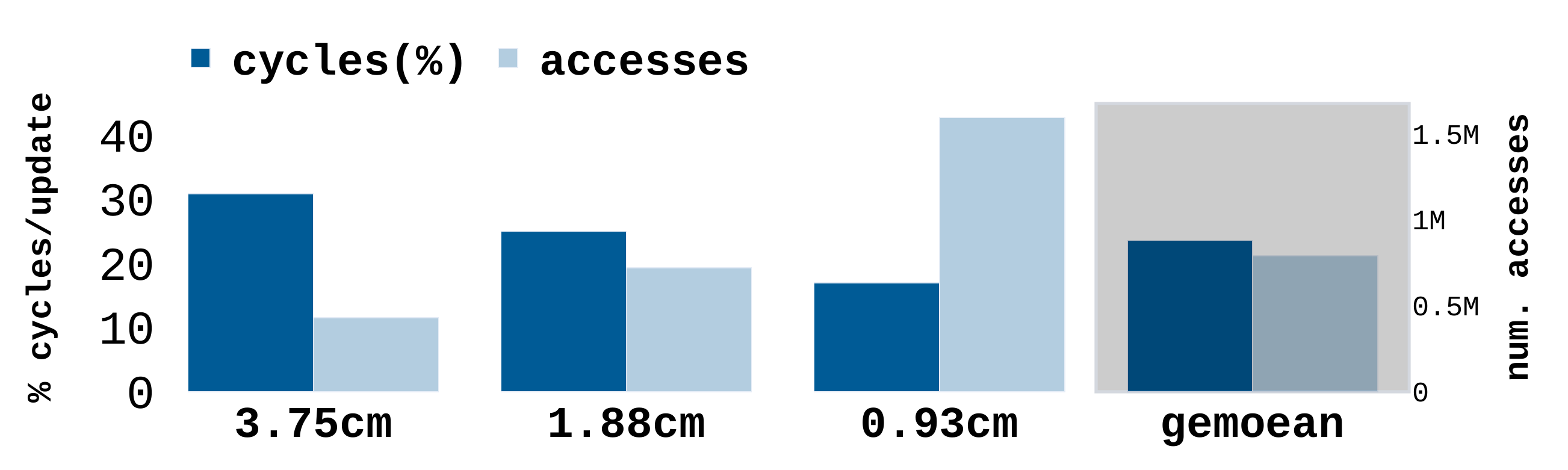}
    \caption{\draftmod{Fraction of time attributed to octree voxel accesses in Supereight~\cite{efficient_largescale_reconstruction}}}
    \label{fig:supereight_access_time}
\end{figure}

\draftmodsec{Figure~\ref{fig:supereight_access_time} shows the number of accesses and the access time as a percentage of the overall scene integration time of Supereight~\cite{efficient_largescale_reconstruction}, which uses an efficient octree implementation for the ICL-NUIM~\cite{iclnuim} living room traj2 dataset. We observe that the average number of accesses at smaller voxel grid sizes is up to $1.5\times10^6$, and the accesses take up to $40\%$ of the scene integration time. In addition, a significant portion of the time taken for generating meshes using ray marching involves (up to $45\%$) voxel access times.}\\

\subsection{Characterizing Key Access Patterns}
\label{sec: locality}
\begin{figure}[htb!]
    \centering
    \includegraphics[width=1.1\linewidth,trim={0.0cm 0.5cm 0.0cm 0cm},clip]{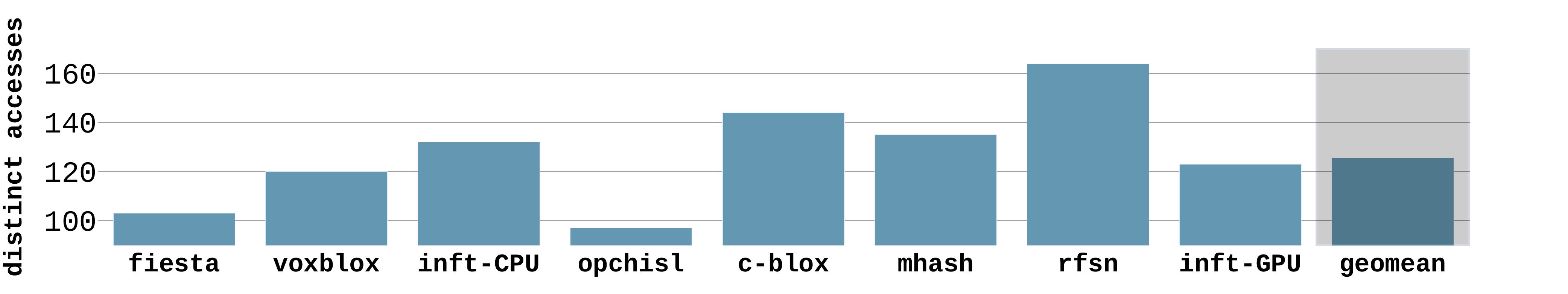}
    \caption{Number of distinct voxels accessed at each map update}
    \label{fig: distinct_accesses}
\end{figure}

Figure \ref{fig: distinct_accesses} shows the number of distinct voxel blocks accessed on average during a map update at a voxel resolution of $0.10m$. On comparing this with the overall number of accesses during each map update step in Figure \ref{fig: accesses and cycles}, we infer that there is massive reuse in accesses to voxel blocks. There are two sources that contribute to this reuse: 
\begin{itemize}[leftmargin=0pt,labelwidth=-10pt]
\item A single update makes repeated accesses to the voxels in a small portion of the large scene. This is because a number of rays cast during raycast intersect the same blocks of voxels when estimating the TSDF. \draftmodfinalremove{In addition, there is a large degree of overlap between regions around the clear radius of update of each voxels when calculating the ESDF(refer to Section~\ref{section: map_update_descr}).} \draftmodfinal{When computing the ESDF, voxels surrounding each point from a point cloud captured by the sensor are updated (refer to Section~\ref{section: map_update_descr}). Since these points (projected from a single depth image) are close by in space, there is a large degree of overlap between update regions of occupied voxels}. As a result, the voxels in these overlapped regions are accessed multiple times during each map update.
\item There is a large overlap in the blocks of voxels accessed in two consecutive updates, as it involves updating the SDF in the same portion of the scene. 

\end{itemize}

\begin{figure}
    \centering
    \includegraphics[width=1\linewidth,trim={0.0cm 0.5cm 0.0cm 0cm},clip]{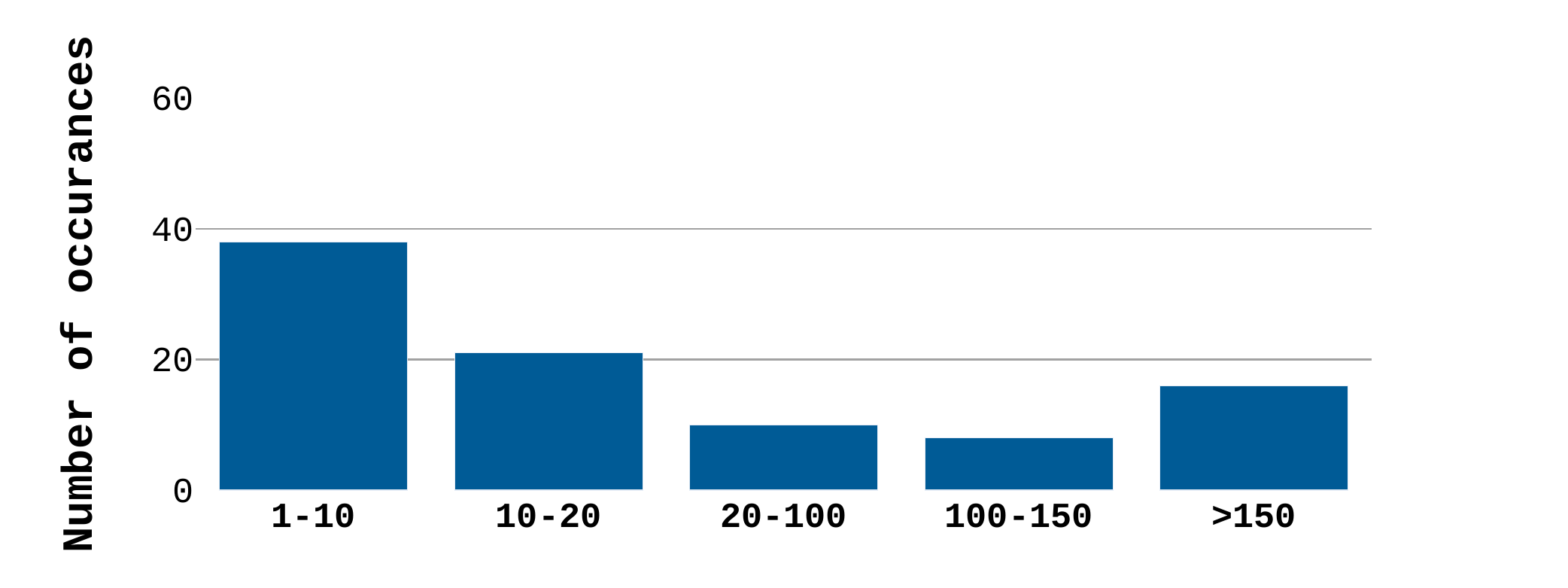}
    \caption{Distribution of the number of map accesses between consecutive accesses to the same key for FIESTA~\cite{fiesta}}
    \label{fig: temporal_locality}
\end{figure}

To empirically verify repetitive key access patterns, we analyze the gaps between accesses to the same key (in terms of map accesses) and count the number of occurrences of each gap for the mapping framework FIESTA \cite{fiesta}. Figure \ref{fig: temporal_locality} shows the histogram of the gap distribution for accesses to the same blocks during a single map update at a voxel resolution of $0.1m$. We observe that a large percentage of gaps between accesses to the same block occur within 150 accesses.

\section{Approach}
\begin{figure}[htb!]
    \centering
    \includegraphics[width=\linewidth]{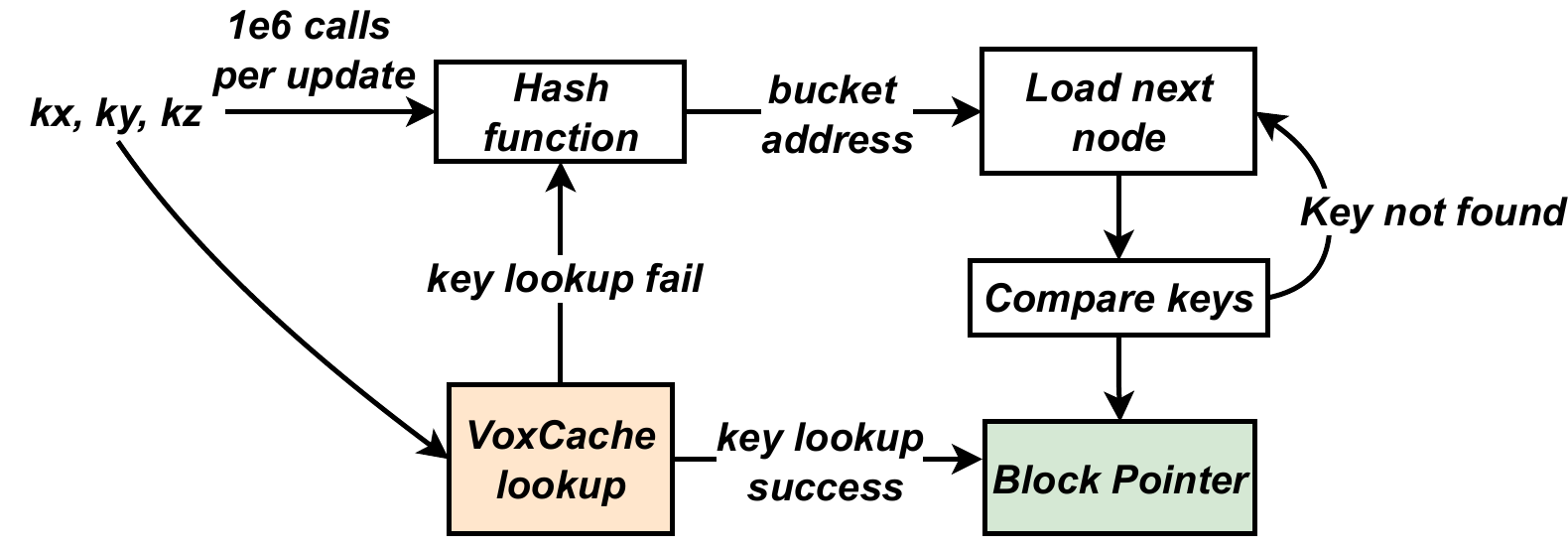}
    \caption{\Xlabel{} lookup path: Each hash table access to the block pointer \draftmodfinalremove{to the}\draftmodfinal{can} be bypassed with a faster lookup }
    \label{fig:lookup_path_high_level}

\end{figure}

Our goal in this work is to enable fast map updates at higher resolutions, which is crucial for many important applications. To this end, we introduce \Xlabel{}, a hardware-software mechanism to enable fast map updates by speeding up voxel data retrieval latencies. The key idea of \Xlabel{} is to leverage temporal locality in accesses to the voxel map by caching the 3D coordinate of the voxel-block + voxel information pair. \draftmodfinalremove{Fast voxel data retrieval is achieved with hardware support for key-indexed lookup of voxel block pointers from the cache.}\draftmodfinal{With hardware support for key-indexed lookup of voxel block pointers from the cache, we enable fast voxel data access.} Figure \ref{fig:lookup_path_high_level} demonstrates the high level mechanism of \Xlabel{}. Voxel block pointers are indexed by its coordinates $kx, ky, kz$ in space. On receiving a request to access voxel data from the block at coordinates $kx,ky,kz$, \Xlabel{} performs a lookup for the corresponding block pointer in the cache and bypasses the costly execution path of \draftmodsec{voxel data access}.
 
We devise a scheme to make use of the existing L1D/L2 cache in CPUs and the L1D cache in GPUs as storage to cache key-block pointer pairs. We introduce a new storage mode which would allow application data to be stored in the cache alongside key-block pointers. We introduce changes to the cache controller to handle two different kinds of requests: virtual addresses and \Xlabel{} accesses. \Xlabel{} accesses handle the lookup/store operations of key-block pointer pairs in the cache. \Xlabel{} provides hardware primitives to store and lookup key-block index. Finally, we expose these primitive operations as ISA instructions accessible in software. 

\subsection{Design Overview}
\label{design overview}
The cache holds two types of data: 1) cached voxel block data, and 2) virtual memory addressable data. We make the following design choices to handle accesses to voxel information and virtual-memory indexed data from the same cache: 

First, we need to define how our key-value data is indexed and placed in the cache. We reserve $m$ ways of an $n$-way set associative cache for voxel block pointer data. Each line holds up to 3 key-block pointer pairs, packed in a contiguous manner. Each key is associated with a pseudoaddress from which the set and tag of the reserved line is derived. The pseudoaddress is computed as:
\begin{equation}
\label{eqn:pseudoaddress}
pseudoaddress = hash\_function(k)\% NR 
\end{equation}
where $NR$ is the number of reserved lines of the outermost reserved level of the cache hierarchy. The pseudoaddress determines which cache line a key-block pointer pair would belong to at each cache level. Note that this definition of the pseudoaddress ensures that keys packed in the same line do not get assigned to different lines in other levels of the cache hierarchy. The position of the key-block pointer pair within a cache line is determined at insert time: If there is an empty slot available, the key-block pointer is inserted into that position. If not, \Xlabel{} replaces the least recently used key-block pointer pair in the line with the new item.

Second, we need to define a replacement policy for the reserved lines to handle writes and evictions in the event of a cache miss or hit. When inserting key-block pointers, a tag and set index is computed for the pseudoaddress. If the tag is not found in the reserved lines in the cache, a cache miss is registered and the request is forwarded to higher levels of the hierarchy. If the tag exists among the reserved lines of the set, the key is compared against the keys in the cache line. The least recently used key-value pair in the line is replaced with the new key-value pair. We implement a writethrough mechanism for all cache levels. On receiving a lookup operation, the cache controller looks up the line corresponding to the set and tag derived from the pseudoaddress of the key. If the tag exists among the reserved section but the key is not found, an invalid value is returned. If the key is found, the corresponding value is returned and the line overwrites the line in the lower levels of the cache hierarchy. In the event where we encounter a cache miss at the outermost reserved section of the cache, we return an invalid value. Unlike a regular cache lookup, we do not forward this request to main memory.

Third, \Xlabel{} introduces \texttt{\Xinstructionprefix{}\_insert}, \texttt{\Xinstructionprefix{}\_remove} and \texttt{\Xinstructionprefix{}\_lookup} instructions to the ISA. \texttt{\Xinstructionprefix{}\_insert} inserts a \Xlabel{} store request, while \texttt{\Xinstructionprefix{}\_lookup} inserts a \Xlabel{} load transaction request into the load store queue. These requests are handled by the cache controller which performs the corresponding insert and lookup operations. In application code, \texttt{\Xinstructionprefix{}\_lookup} is inserted before a voxel block access for lookup. If the lookup is unsuccessful, the application retrieves the voxel block pointer from the \draftmodsec{voxel data structure} after which \texttt{\Xinstructionprefix{}\_insert} is called.

\subsection{\Xlabel{} Inserts and Lookups}

On receiving an insert or lookup or a remove transaction request for voxel block data, the requested key coordinates along with the computed pseudoaddress is inserted into the load-store queue of the core, as demonstrated in Figure~\ref{fig:address_generation_and_issue}. The cache controller picks up these attributes and derives the set index and tag for the cache level from the pseudoaddress. If the corresponding tag does not exist among the reserved lines, a cache miss is registered and the request is forwarded to higher levels of the cache. If the tag exists among the reserved lines of the set, the 16-byte key is compared against the keys in the cache line. For an \texttt{\Xinstructionprefix{}\_insert} operation, the least recently used key-value pair in the line is replaced with the new key-value pair. For a \texttt{\Xinstructionprefix{}\_lookup} operation, if the key is not found, an invalid value is returned. If the key is found, the corresponding value is returned and the line overwrites the line in the lower levels of the cache hierarchy.

\subsection{\Xlabel{}: Example Usage}
Figure~\ref{fig: insert operation} shows an example on how \Xlabel{} is placed in application code to retrieve block pointers. The pseudo code describes the wrapper function to retrieve block pointer corresponding to a requested key $k$ using \Xlabel{}. We first perform a lookup to search for the requested key. If the key is found, the block pointer is returned and we can bypass performing the expensive \draftmodsec{voxel data access}. If we do not find the requested key in the cache, an invalid value is assigned to the block pointer and we do a \draftmodsec{voxel data structure} lookup. After the \draftmodsec{voxel data access}, we insert this key-block pointer pair in the cache for future lookups.

\begin{figure}[!htb]

 \colorbox[RGB]{239,240,241}{
     \begin{minipage}{.9\linewidth}
\begin{algorithmic}

\Function{GetBlkPtr}{$k: key\_coordinates$}
\State $blkptr \gets $\Call{\Xinstructionprefix{}\_lookup}{$k$}
\If{ \Call{is\_invalid}{$blkptr$}}
    \State $blkptr \gets $\Call{ht\_lookup}{$k$}
    \State \Call{\Xinstructionprefix{}\_insert}{$k$,  $blkptr$}
\EndIf
\State \Return $blkptr$
\EndFunction
\end{algorithmic}

\end{minipage}}

\caption{\Xlabel{} lookup algorithm to retrieve the requested block pointer} 
\label{fig: insert operation}
 \vspace{-10pt}

\end{figure}

\section{Detailed Design}
In this section we provide a detailed description of the components of \Xlabel{}, and describe their individual operation.

\subsection{\Xlabel{} Data Indexing and Address Generation}
We design \Xlabel{} to handle key-block pointer data store and load requests similar to regular store and load operation with virtual memory addresses. We describe the changes to the core to handle data accesses to both cached block pointer data and regular virtual memory indexed data.

\subsubsection{Indexing Key-block Pointer Pairs} 

As outlined in Section \ref{design overview}, the cache holds key-block pointer pairs in lines of the cache separate from regular virtual address indexed data. To find the appropriate reserved cache line a key-block pointer pair should belong to, the cache controller would need to derive a set index and tag for a given key at each cache level. To this end, we associate each key with a \Xlabel{} pseudoaddress (Eq. \ref{eqn:pseudoaddress}) from which the set index and the tag can be derived.

Since a single cache line can hold multiple key-block pointer pairs, a naive cache replacement policy evicts and inserts data at a key-block pointer element level. This would require defining a fine grained cache replacement policy which would become exceedingly complicated. To greatly simplify formulating a replacement policy, we make a design choice to index key-block pointer data that enables us to perform evictions and write backs at a cache line level similar to ordinary cache replacement policies. We therefore need to ensure that the cache is able to write back entire lines containing key-block pointer items at a time, without the need to derive locations to individual keys within the line. Therefore it is required that any two keys which derive the same set index and tag at one cache level derive set index and tag equal to each other at all levels of the cache. This criterion ensures that all key-block pointers within a cache line do not derive different set and tag bits at different levels of the cache. A pseudoaddress as defined in Eq. \ref{eqn:pseudoaddress} ensures that any pair of keys packed in one line generate mutually equal set and tag bits at every reserved level of cache.

\begin{figure}[htb!]

    \centering
    \includegraphics[width=\linewidth]{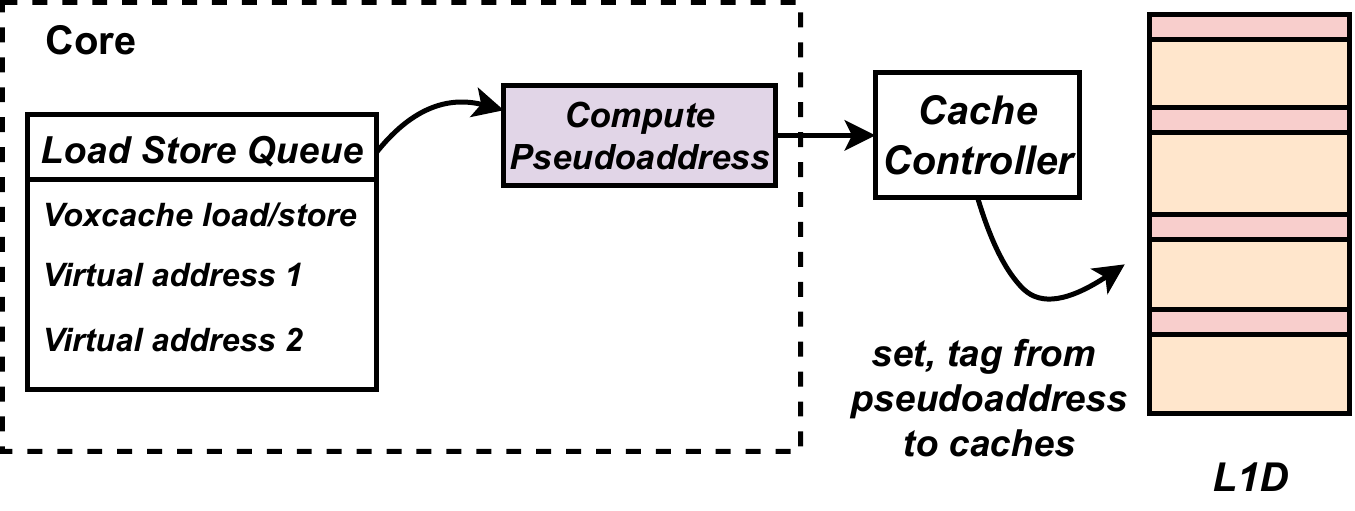}
    \caption{\Xlabel{} issues memory load operations from the load store queue. The red regions of the cache indicate regions where key-block pointer data is stored}
    \label{fig:address_generation_and_issue}

\end{figure}

\subsubsection{Addressing Modes} Under \Xlabel{}, two types of data accesses are to be handled by the core: virtual memory load/stores and key-block pointer load/stores. We augment the core to handle both of these types of loads and stores. \Xlabel{} queues key-block pointer requests to the load store queue alongside regular data indexed by virtual memory as shown in Figure~\ref{fig:address_generation_and_issue}. We categorize the type of access by associating each load/store with one of the two addressing modes: 1) the virtual address mode, which handles memory requests of data indexed by virtual addresses and 2) \Xlabel{} mode, which looks up key-block pointer pairs stored in the cache. Each mode of access follow different cache replacement policies during lookup and insertion. An additional flag in each load store queue entry specifies the addressing mode of our requested data. Note that using the load store queue for \Xlabel{} loads and stores allows leveraging memory disambiguation and store-load queue forwarding techniques supported by the processor leading to better performance.

\subsection{Changes to the Cache Controller/Memory Pipeline}
\Xlabel{} uses on-chip L1D/L2 cache to store key-block pointer pairs. The cache now holds two types of data: 1) key-block pointer data, and 2) regular virtual memory addressable data. We implement additional logic to support lookup and insert operations under \Xlabel{}, and make the following modifications to the cache controller.

\subsubsection{Reserving cache lines in L1D/L2 cache}
\Xlabel{} reserves cache lines in the L1D/L2 caches to store key-block pointer data. Any lookup or insert for a requested key requires \Xlabel{} to perform a search among the reserved lines of the cache. To indicate whether a line is reserved under \Xlabel{} or not, we introduce a 1 bit \texttt{"mode"} flag in the cache line state as shown in Figure~\ref{fig:cache_line_format}.
In an n-way set associative cache, the cache buffer is split into sets of lines with each set containing $n$ cache lines. We implement a hardware primitive, \texttt{reserve\_cache\_lines(m, level)} to mark the first $m$ lines among the $n$ lines per set in the cache for voxel block caching. This primitive invalidates all marked lines of the cache, evicts the data and sets the \texttt{"mode"} flag in the cache state of each line.

\subsubsection{Cache Line Format} 
\begin{figure}[htb!]

\begin{subfigure}{0.5\textwidth}
    \centering
    
    \includegraphics[width=\linewidth]{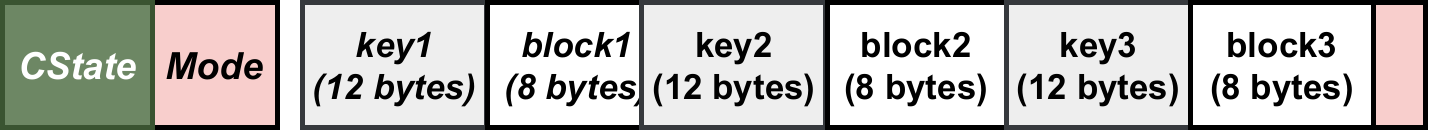}
    \caption{64KB cache line in CPUs: 3 keys, block pointer indices are packed in each reserved line.}
    \label{fig:cache_line_format_cpu}
\end{subfigure}

\begin{subfigure}{0.5\textwidth}
    \centering
    \includegraphics[width=\linewidth]{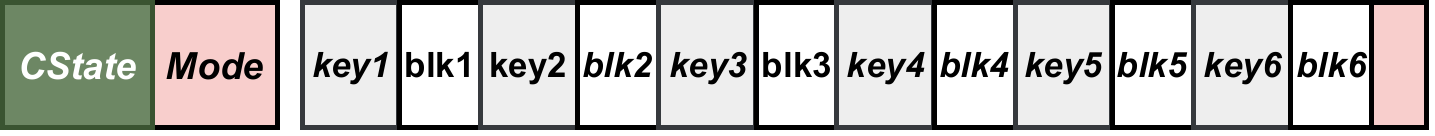}
    \caption{128KB cache line in GPUs: 6 keys, block pointer indices are packed in each reserved line.}
    \label{fig:cache_line_format_gpu}
\end{subfigure}
\caption{\Xlabel{} reserved cache line format for storing keys in CPUs and GPUs. We add an additional flag 'mode' to the cache state.}
\label{fig:cache_line_format}
     \vspace{-10pt}

\end{figure}

We hold multiple key-block pointer pairs in each reserved cache line by contiguously packing key-block pointer bytes of data as depicted in Figure \ref{fig:cache_line_format}. Each key takes up 12 bytes (3D spatial coordinates of integers) and each block pointer takes up 8 bytes of space. For CPU caches with a 64 byte cache line, we pack 3 key-block pointer pairs in a reserved cache line (Figure~\ref{fig:cache_line_format_cpu}) whereas for GPU cache of size 128 kB, we pack 6 key block pointer pairs (Figure~\ref{fig:cache_line_format_gpu}). The position of a key-block pointer pair within the cache line is determined at insert time, where we follow a within-cache line insertion policy to determine the slot it would belong to. We choose an LRU policy to replace exist key-block data within each reserved cache line. To keep track of the least recently used item within the cache line, we introduce additional LRU-bits as part of the cache state.

\subsection{Cache Replacement Policies under \Xlabel{}}
We have separate cache replacement policies for the reserved and unreserved sections of the cache. For virtual memory loads and stores, a normal mode request is sent in the cache controller. Data accesses and evictions of cache lines for normal mode accesses is restricted to the unreserved line sections of the cache. The replacement policy for normal mode requests is identical to the cache replacement policy of the processor, with the constraint that write back and eviction operations only modify unreserved lines of the cache. We effectively trade off the number of lines of cache available to the process for faster accesses to the key-block pointer data. For reserved lines under \Xlabel{}, we need to define $2$ cache replacement policies: \emph{Entire-line replacement policy} and
 \emph{Within-line item replacement policy}.

For key-block pointer loads and stores, the cache controller is restricted to only lookup the \Xlabel{} reserved lines in the cache. We now describe the entire-line level and within-line replacement policy in detail. The cache is said to have registered an entire-line-read-HIT when a \Xlabel{} load request for a key finds the corresponding set and tag among the reserved cache lines during lookup. If the requested key exists among the packed keys in the cache line, we register a within-line-read-HIT, otherwise we register a within-line-read-MISS. An entire-line-read-MISS is registered when the set and tag derived from the requested key is not found in the cache. 

An entire-line-write-HIT is registered when a \Xlabel{} store request for a key finds its derived set and tag in the reserved lines of the cache. A within-line-write-HIT is when the key to be overwritten exists in the cache line. Otherwise a within-line-write-MISS is registered. A entire-line-write-MISS is registered when a \Xlabel{} store for a key does not find find its tag bits in the set in the reserved section of the cache.

\subsubsection{Entire Cache Line Replacement Policy}

\Xlabel{} reserves cache lines in the L1D/L2 levels of the cache of CPUs, and in the L1D buffer of GPUs. Tables \ref{table:cpurepl} and \ref{table:gpurepl} outline the function of the cache controller in each possible scenario for CPUs and GPUs respectively. 

\begin{figure}[htb!]
    \centering
    \includegraphics[width=0.55\linewidth, trim={0 0cm 0 -0.5cm},clip]{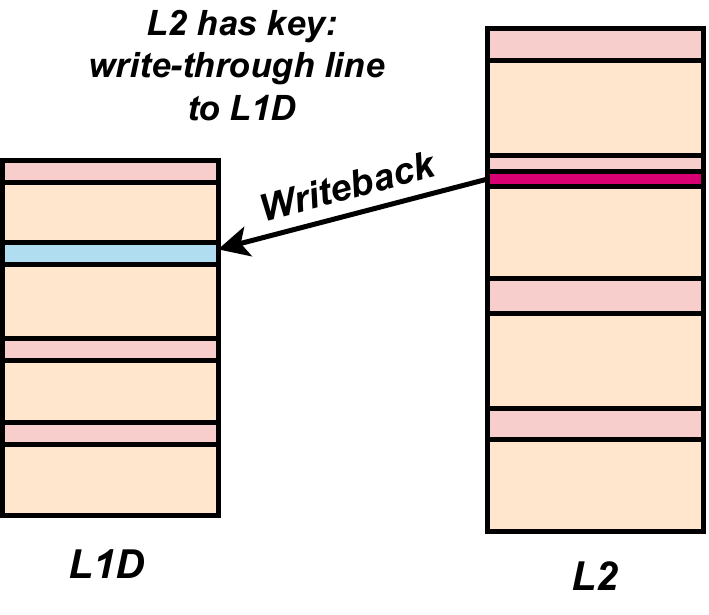}
    \caption{The line corresponding to the key found on L2 cache is written through to L1D }
    \label{fig: l2 writethrough}
\end{figure}

For CPUs, on encountering an entire-line-read-MISS on L1D, the read request is forwarded to L2 cache. If we encounter an entire-line-read-MISS on L2, we return an invalid value. On encountering an entire-line-read-HIT in L1D or L2, The lookup then proceeds to look for the requested key within the line. If we encounter the entire-line-read-HIT at the L2 level, we implement a write back operation which copies the corresponding line into the L1D cache, as shown in Figure~\ref{fig: l2 writethrough}. For an entire-line-write-MISS in L1D, we forward the store request onto L2. On encountering an entire-line-write-HIT, we evict the least recently used reserved cache line in the set and insert the new key-block pointer data. An entire-line-write-HIT would invoke the within line replacement policy.

For GPUs, an entire-line-read-MISS from the L1D cache returns an invalid block pointer. We avoid forwarding \Xlabel{} requests to the L2 level for GPUs. An entire line read-HIT would search within the line for the requested key. On encountering an entire-line-write-MISS on L1D, we evict the least recently used line in the set and insert the new key-block pointer in the line. An entire-line-write-HIT would invoke the within line replacement policy.

\begin{table}[htb!]
  \centering
  \caption{Read-Write replacement policy for \Xlabel{} requests for a CPU}
  \label{table:cpurepl}
  \begin{tabular}{|l|l|l|}
    \hline
    & \textbf{L1D} & \textbf{L2}\\
    \hline
    \textbf{Read hit} & Lookup within line & Writeback \\
    \hline
    \textbf{Read miss} & Forward to L2 & Return invalid  \\
    \hline
    \textbf{Write hit} &  Overwrite within line  & Modified by writethrough  \\
    \hline
    \textbf{Write miss} & Evict LRU line; Insert & Modified by writethrough \\
    \hline
  \end{tabular}
\end{table}

\begin{table}[htb!]
  \centering
  \caption{Cache line replacement policy for \Xlabel{} requests on a GPU}
  \label{table:gpurepl}
  \begin{tabular}{|l|l|}
    \hline
    & \textbf{L1D}\\
    \hline
    \textbf{Read hit} & Lookup within line \\
    \hline
    \textbf{Read miss} & Return invalid \\
    \hline
    \textbf{Write hit} & Overwrite block pointer \\
    \hline
    \textbf{Write miss} & Evict; Write block pointer  \\
    \hline
  \end{tabular}
\end{table}

\subsubsection{Within Cache Line Items Replacement Policy}
A reserved cache line holds multiple key-block pointer items. During loads and stores, we replace the least recently used key-block pointer item present in the cache line. We make use of the within cache line LRU-bits in the cache state to find the least recently used key-block pointer pair within the cache line. These LRU-bits per line are updated each time an insert or lookup operation is performed. Table \ref{table:withinline} describes the within cache line update policies in each scenario.

At load time, if the requested key exists in the cache line, the LRU-bits for the cache line are updated and the block pointer corresponding to the key is returned. If the requested key does not exist, an invalid value is returned. At insert time, a modified line along with the within-line LRU bits is written through to all the levels of the cache. 

\begin{table}[htb!]
  \centering
  \caption{Within line cache replacement policy}
  \label{table:withinline}
  \begin{tabular}{|l|l|}
    \hline
    & \textbf{Within Line Operation}\\
    \hline
    \textbf{Within-line read hit} & Return data, update LRU bits \\
    \hline
    \textbf{Within-line read miss} & Return invalid value \\
    \hline
    \textbf{Within-line write hit} & Update LRU bits \\
    \hline
    \textbf{Within-line write miss} & Overwrite; Update LRU bits  \\
    \hline
  \end{tabular}
\end{table}

\subsection{Software Interface: ISA Instructions}
\Xlabel{} exposes $5$ primitive operations as instructions to the ISA. These instructions control the configuration of the cache and perform accesses to key-block pointer pairs stored in the cache. A summary of their functionality is shown in Table \ref{table: isa}. The \texttt{reserve\_cache\_lines} and \texttt{unreserve\_lines} instructions are used to configure and modify cache state to toggle between \Xlabel{} mode and regular addressability mode. Instructions \texttt{\Xinstructionprefix{}\_lookup}, \texttt{\Xinstructionprefix{}\_insert} and \texttt{\Xinstructionprefix{}\_remove} handle key-block pointer data load, store, or remove from the cache respectively.

\begin{table}[htb!]
  \centering
  \caption{\Xlabel{} ISA Instructions}
  \label{table: isa}
  \begin{tabular}{|l|l|}
    \hline
    \textbf{ISA Instruction} & \textbf{Functionality}\\
    \hline
    \texttt{reserve\_cache\_lines}  & Invalidates first \texttt{ways}\\\texttt{ways, lvl} & lines of all sets of cache in\\ & \texttt{lvl} of the memory hierarchy  \\
    \hline
    \texttt{\Xinstructionprefix{}\_lookup state, } & Takes the 16 bytes of key \\\texttt{rd1, rd2, r1} & from rd1, rd2 registers as inputs, \\ & outputs the value to \texttt{r1}, if found.
    \\
    \hline
    \texttt{\Xinstructionprefix{}\_remove status, } & Takes the 16 bytes of the\\ \texttt{rd1, rd2} & key entry from the rd1, \\ & rd2 and inserts invalid entry\\ & as its value.  \\
    \hline
    \texttt{\Xinstructionprefix{}\_insert status, } & Takes the 16 byte key entry \\ \texttt{rd1, rd2, r1} &  from rd1,rd2 and inserts \\ & the element into cache  \\
    \hline
    \texttt{unreserve\_lines} & Invalidates all \Xlabel{} reserved\\ &  lines and changes mode of each\\ & line back to addressability mode \\
    \hline
  \end{tabular}
\end{table}

\begin{itemize}
    \item \texttt{reserve\_cache\_lines(m, l)}: Communicates to the cache controller to invalidate the first \texttt{m} lines of all sets of the cache at level \texttt{l}, and set the reserved mode flag for each of the \texttt{m} lines. 
    \item \texttt{\Xinstructionprefix{}\_lookup} : Takes the key input from registers \texttt{rd1}, \texttt{rd2} and inserts  returns the value into output register \texttt{r1}, if found. If the corresponding value is not found, an invalid value is assigned to \texttt{r1}
    \item \texttt{\Xinstructionprefix{}\_insert}: Derives the key from input registers \texttt{rd1}, \texttt{rd2} and value from \texttt{r1} and inserts a \Xlabel{} store request in the load store queue of the core, containing the pseudoaddress and the key coordinates. 
    \item \texttt{\Xinstructionprefix{}\_remove}: Derives the key input from registers \texttt{rd1}, \texttt{rd2} and performs an insert operation with an invalid value.
    \item \texttt{unreserve\_lines}: Returns to general mode of execution by invalidating all reserved lines in each set in the cache, and un-setting the reserved mode flag in the cache state of each line.
\end{itemize}

\subsection{Example: \Xlabel{} Lookup Operation Walkthrough}

In CPUs, load and store instructions are handled by virtual memory addresses produced by the address generation unit. In GPUs, each thread in a warp would issue load and store operations to global memory space by inserting requests to a per-SIMT memory access queue of the load-store unit. We insert \Xlabel{} store and load operations in the respective load store queues of CPUs and GPUs. Here, we describe the execution of a \Xlabel{} load operation by the cache controller for CPUs.

\begin{figure}[htb!]

\begin{subfigure}{0.5\textwidth}
    \centering
    \includegraphics[width=.95\linewidth, trim={0cm 0cm 0 0cm},clip]{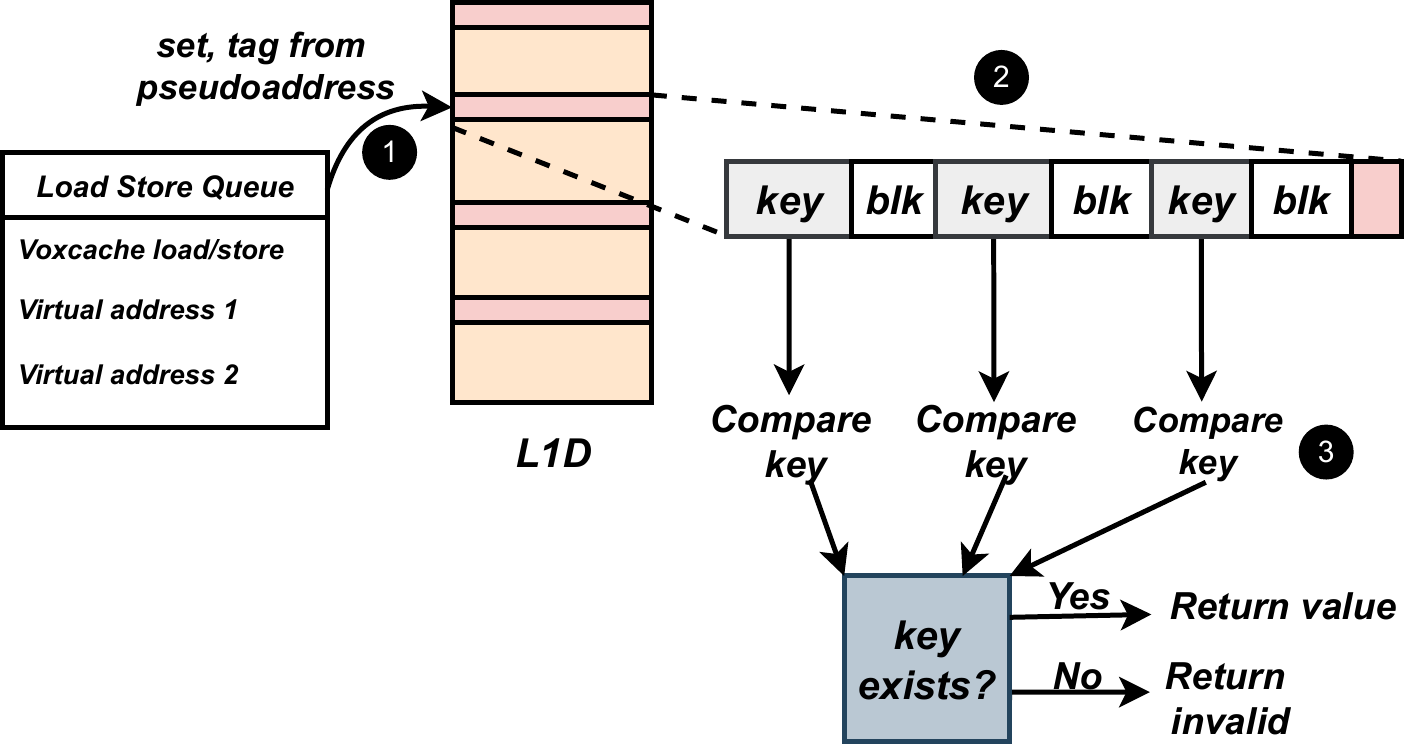}
    \caption{A \Xlabel{} key lookup from the L1D cache.}
    \label{fig:l1_lookup}
\end{subfigure}

\begin{subfigure}{0.5\textwidth}
    \centering
    \includegraphics[width=1.02\linewidth, trim={0.2cm 0cm 0 -0.5cm},clip]{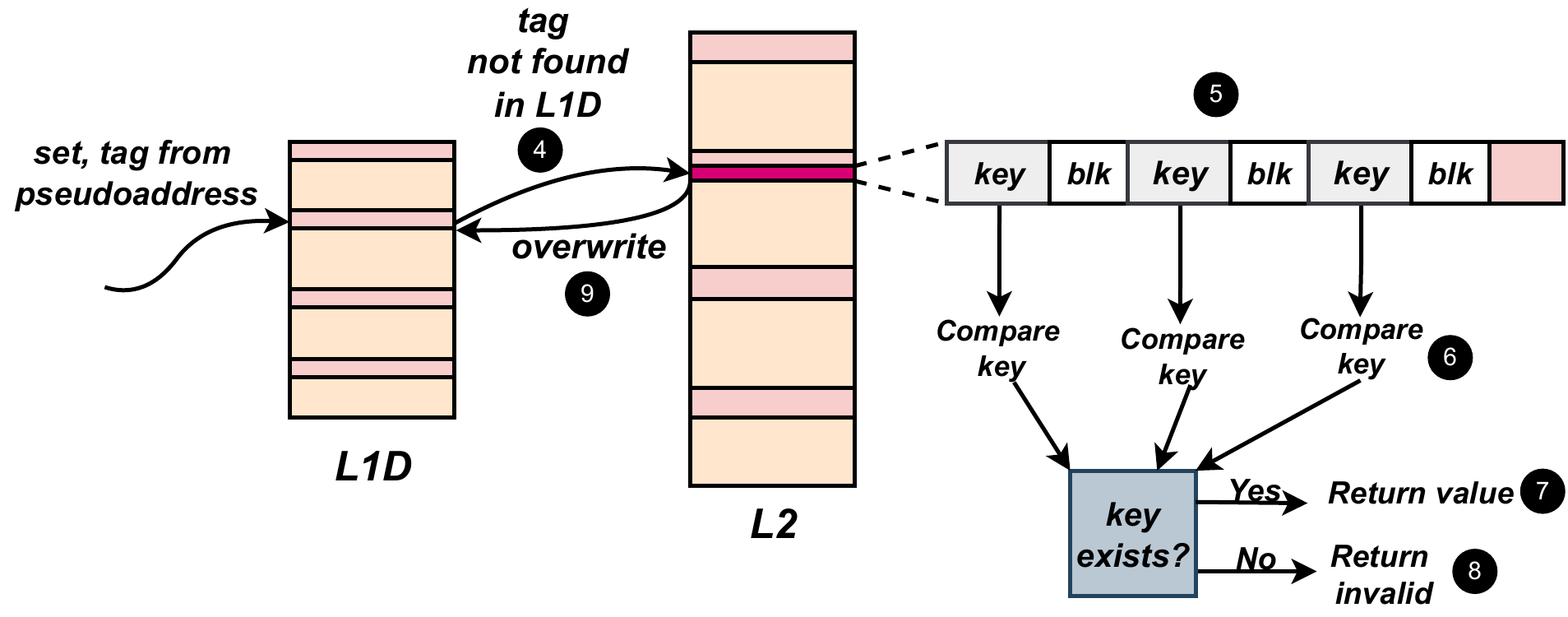}
    \caption{A \Xlabel{} key lookup from the cache: lookup forwarded to L2 following a miss form L1D.}
    \label{fig:l2_lookup}
\end{subfigure}
\label{fig:l1l2_lookup}
\caption{A \Xlabel{} key lookup from the cache. The red regions of the cache represent reserved sections of the cache.}

\end{figure}

Figure \ref{fig:l1_lookup} and \ref{fig:l2_lookup} describe a \Xlabel{} lookup where we reserve 1 way out of 4 and 2 ways out of 8 per set in L1D and L2 respectively, and the key is present at the L2 level. \texttt{\Xinstructionprefix{}\_lookup} inserts a \Xlabel{} load request in the load store queue respectively \circled{1}. The core computes the pseudoaddress of the key from which the cache controller derives the set index and tag from this pseudoaddress for L1D cache \circled{2}. The set and tag derived from the pseudoaddress for L1D are used to find the corresponding line to look up. If the corresponding tag exists in the cache, it checks the cache line for the required key~\circled{3}. If the key is found, the corresponding value is overwritten by the new block pointer.

If the tag is not found at the L1D level, the request is forwarded \circled{4} to L2. The set and tag are now derived for L2. If the corresponding tag exists here, the lookup now proceeds to compare against the keys within the line \circled{5}. Each key within the line is compared with the requested key \circled{6}. If found, this value is returned \circled{7} and line replaces lower level lines in the hierarchy \circled{8}. Otherwise, an invalid value is returned \circled{9}.

\section{Methodology}
We model and evaluate \Xlabel{} on CPUs using the sniper simulator~\cite{snipersim}. Table~\ref{table: cpusim} lists the specifications used. We reserve 1 way per set out of a 4-way set associative cache in L1D, and 2 ways per set out of an 8-way associative L2 cache for \Xlabel{}. For GPU workloads, we use GPGPUSim~\cite{gpgpu-sim} to model a GPU with specifications mentioned in Table~\ref{table: gpusim}. We allocate 1 way out of the 4 ways per set in L1D for \Xlabel{}.

\begin{table}[h!]
  \centering
  \caption{Simulated system configuration: CPU}
  \label{table: cpusim}
  \begin{tabular}{|l|}
    \hline
    \hline
     \textbf{CPU} 3.6GHz Cascade-lake-like, OOO 4-wide dispatch window,\\ 128-entry ROB; 32 entry LSQ \\
    \hline
     \textbf{L1D + L1I Cache} 32KB, 4 way LRU, 1 cycle; 64 Byte line; \\ MSHR size: 10; stride prefetcher \\
    \hline
     \textbf{L2 Cache}  256KB, 8 way LRU, 4 cycle; 64 Byte line; \\ MSHR size: 10; stride prefetcher \\
    \hline
     \textbf{L3 Cache} 1MB, 16 way LRU, 20 cycle; 64 Byte line; \\ MSHR size: 64; stride prefetcher \\
    \hline
     \textbf{DRAM} 2-channel; 16-bank; open-row policy, 4\draftmod{GB} DDR4\\
    \hline
  \end{tabular}

\end{table}

\begin{table}[h!]

  \centering
  \caption{Simulated system configuration: GPU}
  \label{table: gpusim}
  \begin{tabular}{|l|}
    \hline
    \hline
     \textbf{Shader core} 1.4GHz; 2 schedulers per SM \\
    \hline
     \textbf{SM Resources} 32768 Registers, 32KB Shared memory,\\ 128KB L1D, 4 ways \\
    \hline
     \textbf{DRAM} 2-channel; 16-bank; open-row policy, 4\draftmod{GB} DDR4\\
    \hline
  \end{tabular}

\end{table}

\subsection{Workloads and Datasets}
Outlined in Table \ref{table: cpu workload parameters} and Table \ref{table: gpu workload parameters} are the workloads and datasets used to run our experiments for CPUs and GPUs respectively. \draftmod{The EuRoC Machine hall dataset~\cite{euroc} is an RGB-D indoor dataset consisting of depth maps and images captured from motion across a large cluttered indoor industrial environment. The KITTI~\cite{kitti} sequence is an outdoor RGB-D dataset captured on driving a vehicle across a neighborhood. These datasets consist of data captured from motion over a relatively large area of space, where a voxel hashing implementation is required to keep track of the environment’s map in memory.}

\textbf{CPU workloads}:
\begin{itemize}
    \item  \texttt{\textbf{voxblox}}: Voxblox~\cite{voxblox} is a mapping framework for constructing truncated signed distance fields (TSDF) and Euclidean signed distance fields (ESDF) for MAV planning applications.
    \item \texttt{\textbf{inft-CPU}}: InfiniTAM~\cite{infinitam} is a popular framework for large-scale 3D reconstruction with loop closure, built on top of KinectFusion~\cite{kinectfusion}.
    \item  \texttt{\textbf{fiesta}}: FIESTA~\cite{fiesta} is an efficient mapping system to compute the Euclidean Signed Distance Field (ESDF) map on the fly. The constructed map is used for online motion planning for aerial robotics. 
    \item \texttt{\textbf{opchisl}}: OpenChisel~\cite{openchisel} is a dense 3D reconstruction framework for mapping and localization, targeted for Google Tango mobile device.
    \item \texttt{\textbf{c-blox}}: c-blox~\cite{cblox} is a TSDF mapping library that is built on top of voxblox, designed for consistancy for scalability in very large scale mapping.
    \item \draftmodsec{\texttt{\textbf{supereight}}: supereight~\cite{efficient_largescale_reconstruction} is a TSDF mapping library which uses an efficient octree implementation to store voxel data}.

\end{itemize}

\textbf{GPU workloads}:
\begin{itemize}
    \item \texttt{\textbf{rfsn}} : Refusion~\cite{refusion} is a GPU based fast 3D reconstruction and meshing framework that is designed to be robust to dynamic environments.
    \item \texttt{\textbf{mhash}}: Mesh hashing~\cite{mesh_hashing} proposes a framework to incrementally generate, update meshes online on the GPU from spatial hashed data structure.
    \item \texttt{\textbf{inft-GPU}}: GPU version of infiniTAM~\cite{infinitam}, for large scale mobile 3D reconstruction.
\end{itemize}

\begin{table}[h!]

  \centering
  \caption{CPU Workload configuration}

  \label{table: cpu workload parameters}
  \begin{tabular}{|l|l|}
    \hline
    \textbf{Workload}   & \textbf{Dataset}   \\
    \hline
    \hline
     Voxblox~\cite{voxblox} & KITTI Sequence 0027~\cite{kitti} \\
    \hline
     OpenChisel~\cite{openchisel}   & EuRoC Machine Hall Sequence 5~\cite{euroc}  \\
    \hline
     FIESTA~\cite{fiesta}       & KITTI Sequence 0027~\cite{kitti} \\
    \hline
     InfiniTAM~\cite{infinitam} & \draftmod{EuRoC Machine Hall Sequence 5~\cite{euroc}} \\
    \hline
     c-blox~\cite{cblox}        & KITTI Sequence 0027~\cite{kitti} \\
    \hline
     supereight~\cite{efficient_largescale_reconstruction}        & ICL NUIM dataset~\cite{iclnuim} \\
    \hline

  \end{tabular}
\end{table}

\begin{table}[h!]
  \centering
  \caption{GPU workload configuration}
  \vspace{-10pt}
  \label{table: gpu workload parameters}
  \begin{tabular}{|l|l|}
    \hline
    \textbf{Workload} & \textbf{Dataset}   \\
    \hline
    \hline
     ReFusion~\cite{refusion}      & KITTI Sequence 0027~\cite{kitti} \\
    \hline
     InfiniTAM~\cite{infinitam}     & EuRoC Machine Hall Sequence 5~\cite{euroc}  \\
    \hline
     MeshHashing~\cite{mesh_hashing}  & KITTI Sequence 0027~\cite{kitti} \\
    \hline
  \end{tabular}
\end{table}

\section{Evaluation}
\subsection{Speedup on Map Update}
\begin{figure}[htb!]

\begin{subfigure}{0.5\textwidth}
    \centering
    \includegraphics[width=1.\linewidth,trim={0.1cm 0.5cm 1.7cm 0cm},clip]{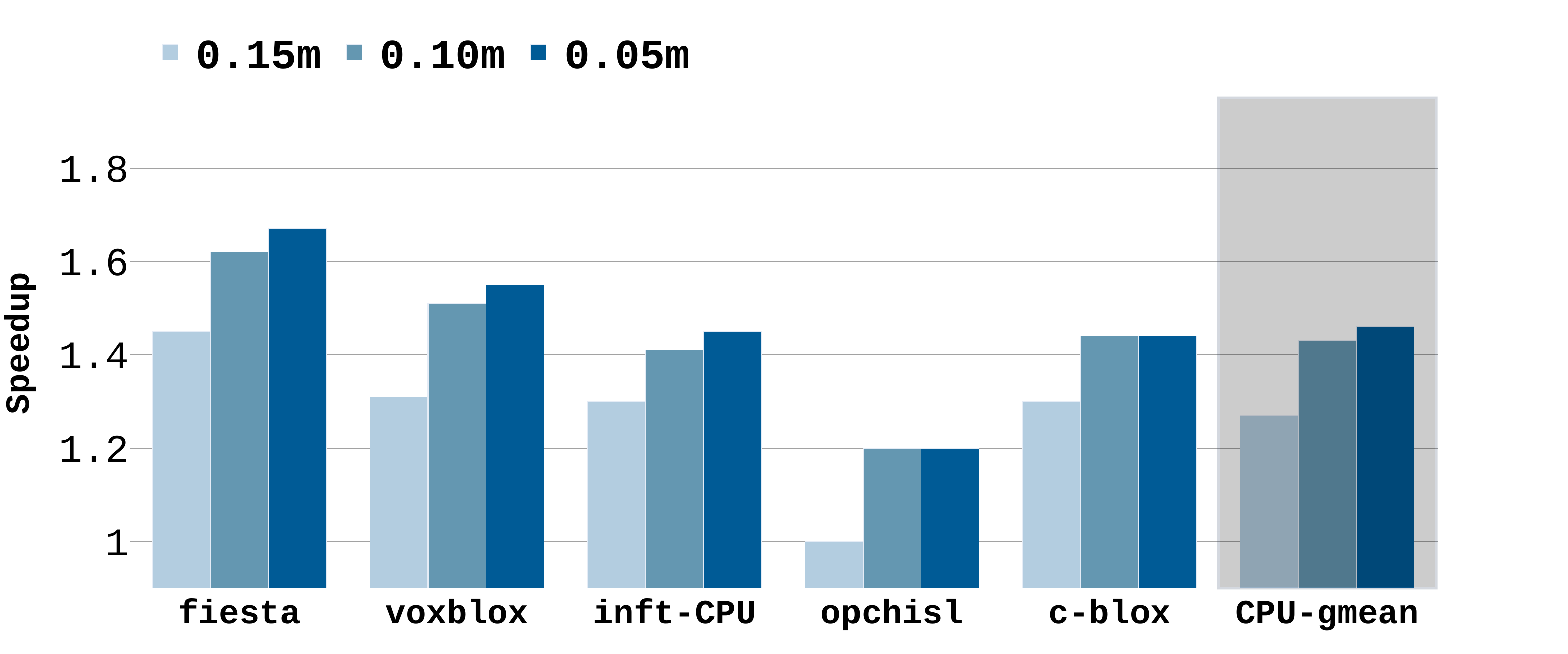}
    \caption{Speedups on CPU workload}
\end{subfigure}

\begin{subfigure}{0.5\textwidth}
    \centering
    \includegraphics[width=1.\linewidth,trim={0.cm 0.5cm 0 0.5cm},clip]{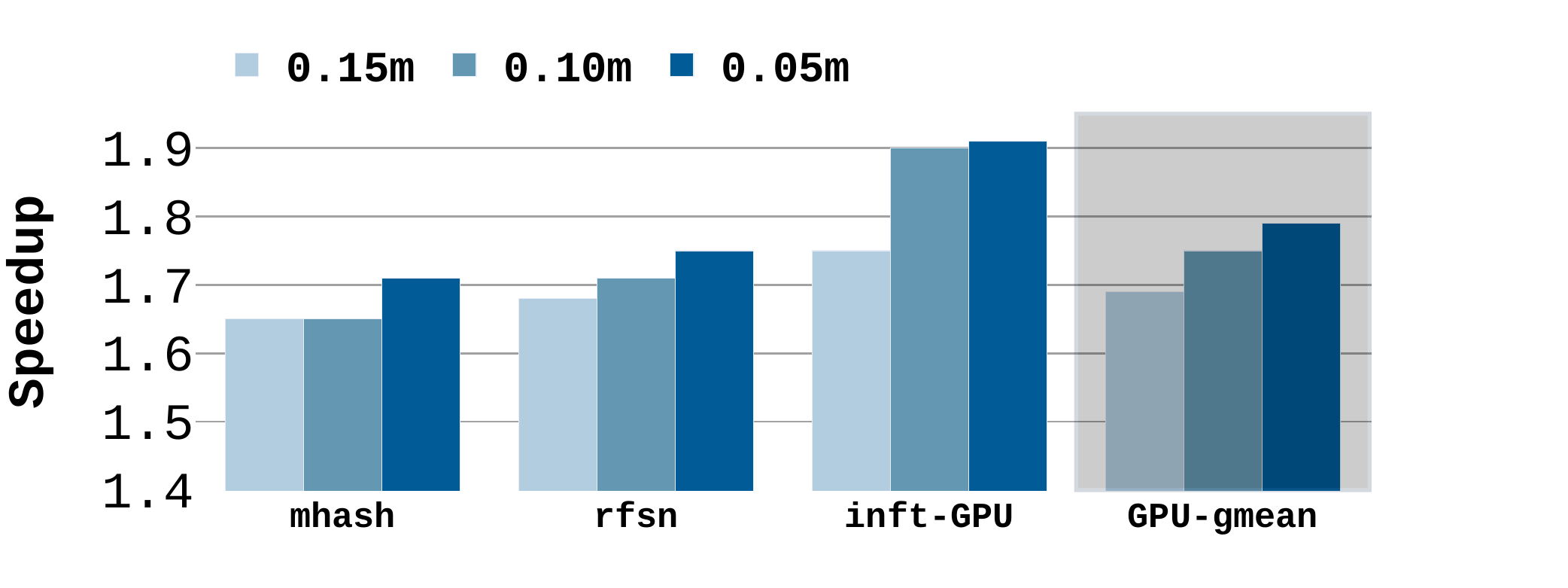}
    \caption{Speedups on GPU workload}
     \vspace{-5pt}

\end{subfigure}
\caption{Change in speedup on varying the voxel grid size}
\label{fig: speedup resolution}

\end{figure}

We measure the speedup on map update at different resolutions (at different voxel grid sizes). Depicted in Figure \ref{fig: speedup resolution} are the speedups obtained at resolutions $5cm$, $10cm$ and $15cm$, where we observe average an speedup of $1.37X$ on CPU and $1.74X$ on GPU. We observe that we get higher speedups at higher resolutions. On average, at the highest resolution ($5cm$ voxel grid size), we achieve a speedup of \draftmod{$1.45X$} on CPU and $1.78X$ on GPU. We conclude that \Xlabel{} can effectively improve map update times at different resolutions.

\subsection{Energy Analysis}
\begin{figure}[htb!]

\begin{subfigure}{0.5\textwidth}
    \centering
    \includegraphics[width=1.\linewidth,trim={0.1cm 0.4cm 0 0cm},clip]{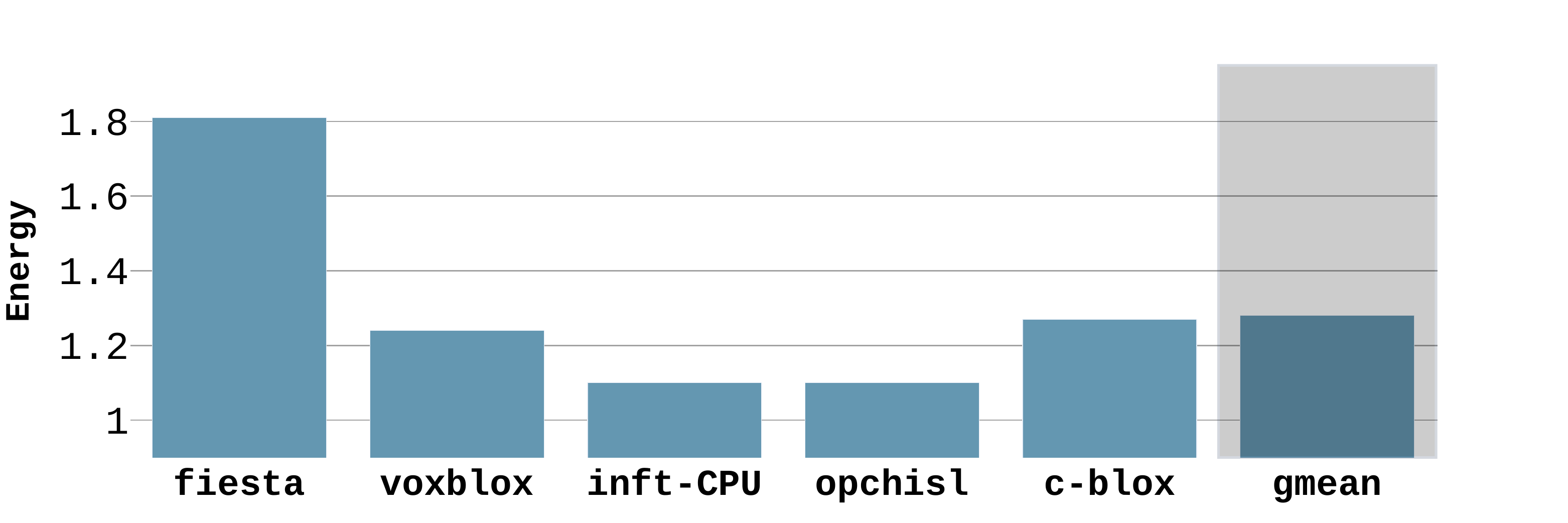}
    \caption{Energy consumed on CPU workloads}
    \label{fig: cpu energy}
\end{subfigure}
\begin{subfigure}{0.5\textwidth}
    \centering
    \includegraphics[width=1.\linewidth,trim={0.1cm 0.4cm 0 0cm},clip]{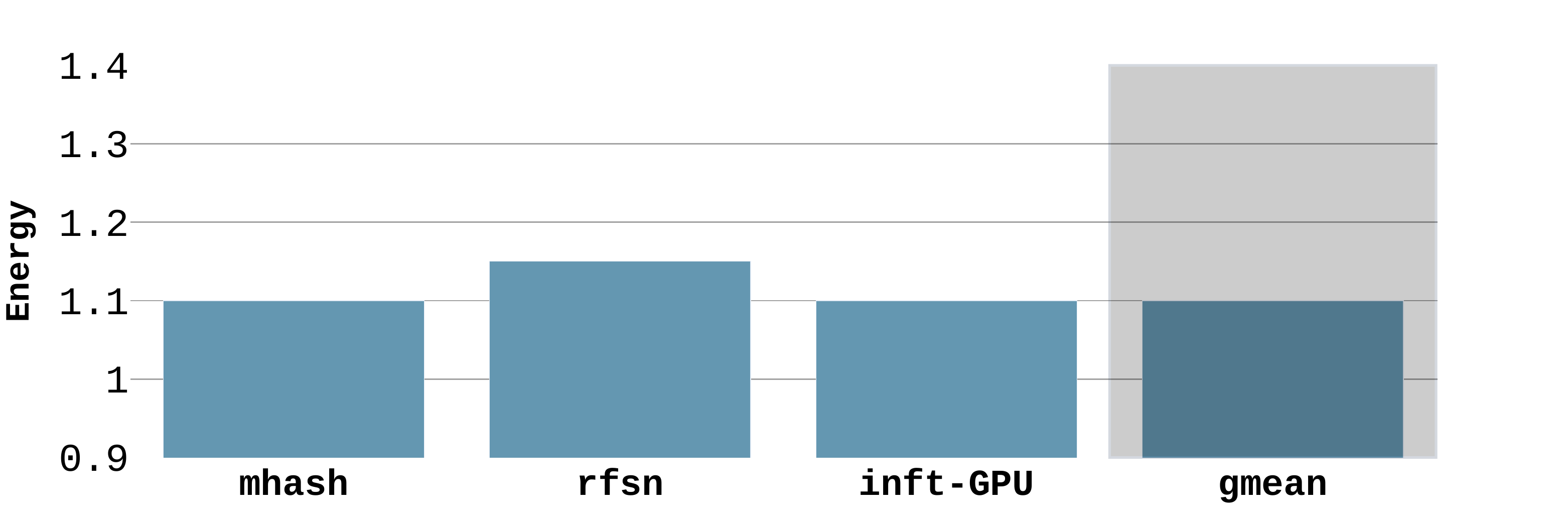}
    \caption{Energy consumed on GPU workloads}
    \label{fig: gpu energy}
\end{subfigure}

\caption{Normalized energy consumed per map update}

\end{figure}

Figure \ref{fig: cpu energy} and \ref{fig: gpu energy} shows the energy savings of \Xlabel{} normalized to baseline on CPU  and GPU workloads respectively at a voxel grid size of 0.05m. \Xlabel{} on average requires 22\% less energy (up to 44\%) on CPU and 9.5\% (up to 14\%) on GPU respectively. \Xlabel{} consumes less energy for mapping as a result of fewer accesses to DRAM by significantly reducing the number of hash resolutions required, which requires higher energy consumption and leads to longer runtime.

\subsection{Comparison to Using a Fully Associative Buffer}
We consider the case where we use an idealized fully associative buffer with an LRU replacement policy to store and lookup block pointers using keys and compare the cache miss rates with that of \Xlabel{}. Note that we count an access to \Xlabel{} as a hit if the requested voxel block is found in reserved cache lines at any level of the cache hierarchy.

\begin{figure}[htb!]
\begin{subfigure}{0.5\textwidth}
    \centering
    \includegraphics[width=1.1\linewidth,trim={0 1cm 0 0cm},clip]{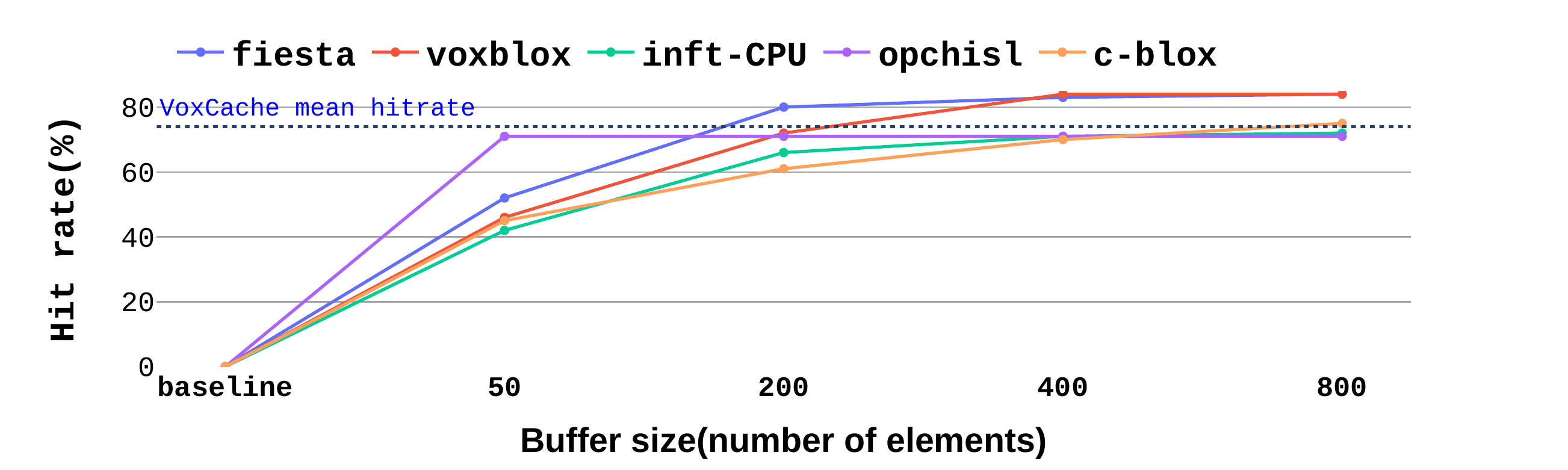}
    \caption{Hit rate vs buffer size observed for CPU workloads}
    \label{fig: cpu hit rates}
\end{subfigure}

\begin{subfigure}{0.5\textwidth}
    \centering
    \includegraphics[width=0.8\linewidth,trim={0 0cm 0 0cm},clip]{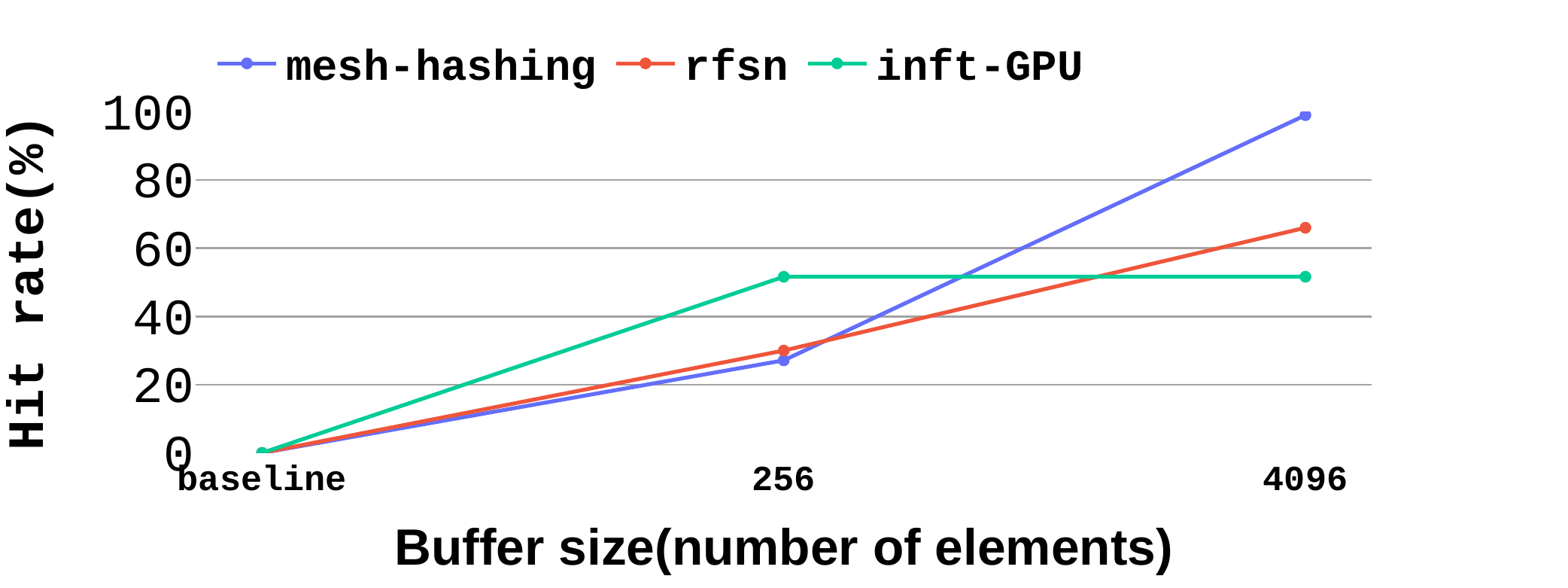}
    \caption{Hit rate vs buffer size observed for GPU workloads}
    \label{fig: gpu hit rates}
\end{subfigure}
    \caption{Hit rates vs fully associative buffer size (number of key-block pointer pairs held)}
 \vspace{-10pt}

\end{figure}

Figure \ref{fig: cpu hit rates} shows the hit rate on using a fully associative buffer for lookups on CPU workloads, using workload configurations as mentioned in Table \ref{table: cpu workload parameters}. We observe that the number of cache hits we receive saturates for a buffer size of around $400$ elements. We also  observe that the performance benefit we achieve with a fully associative buffer which we achieve cache hit rates which compare closely to that of using \Xlabel{}. The hit rates for GPU workloads are as shown in Figure \ref{fig: gpu hit rates}. A cache hit is said to have registered when all threads of the warp encounter a cache hit.

\begin{figure}[htb!]

\begin{subfigure}{0.5\textwidth}
    \centering
    \includegraphics[width=\linewidth,trim={0 0cm 0 0cm},clip]{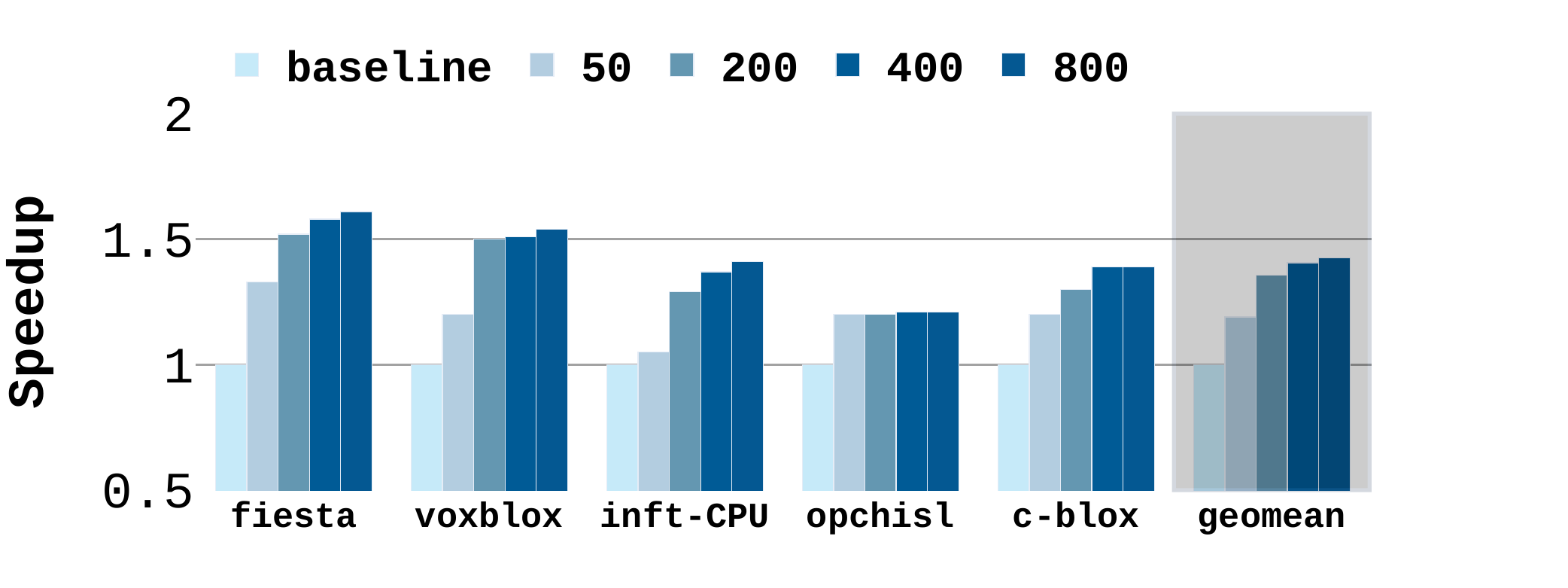}
    \caption{Speedup on GPU applications}
    \label{fig: cpu buffer size}
\end{subfigure}

\begin{subfigure}{0.5\textwidth}
    \centering
    \includegraphics[width=0.9\linewidth,trim={0 0cm 0 0cm},clip]{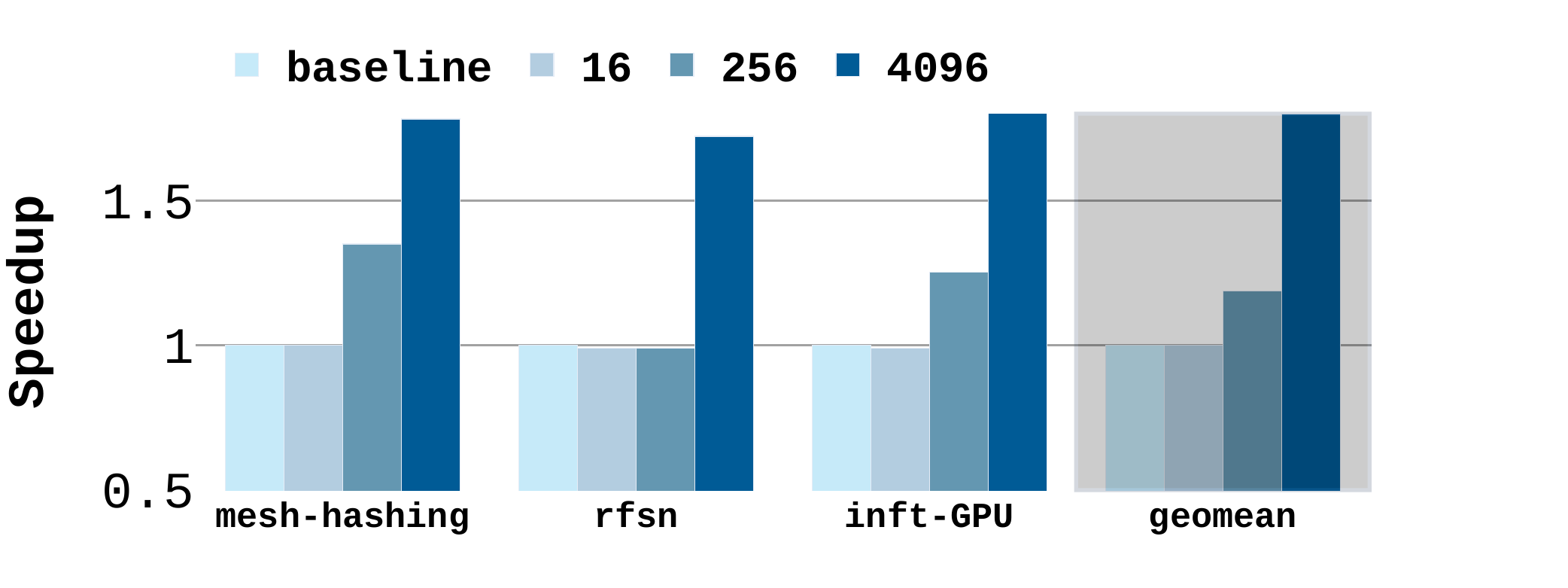}
    \caption{Speedup on GPU applications}
    \label{fig: gpu buffer size}
\end{subfigure}

\caption{Speedup observed on varying the fully associative buffer size}
 \vspace{-15pt}

\end{figure}

With an access latency set to the same as the lookup times for L1D caches, the speedup achieved for a fully assocative buffer storking key-value pointers is shown in Figure \ref{fig: cpu buffer size} for CPU workloads, and in Figure \ref{fig: gpu buffer size} for GPU workloads respectively. We achieve a speedup of $1.9\times$ on InfiniTAM-GPU, $1.78\times$ on mesh-hashing and $1.73\times$ on ReFusion for a buffer size of 4096 elements.

\subsection{Comparison to Hash Table Accelerator}
\label{sec:comparison_hta}
\begin{figure}[htb!]
    \centering
    \includegraphics[width=1.1\linewidth,trim={0 0.3cm 0 0.5cm},clip]{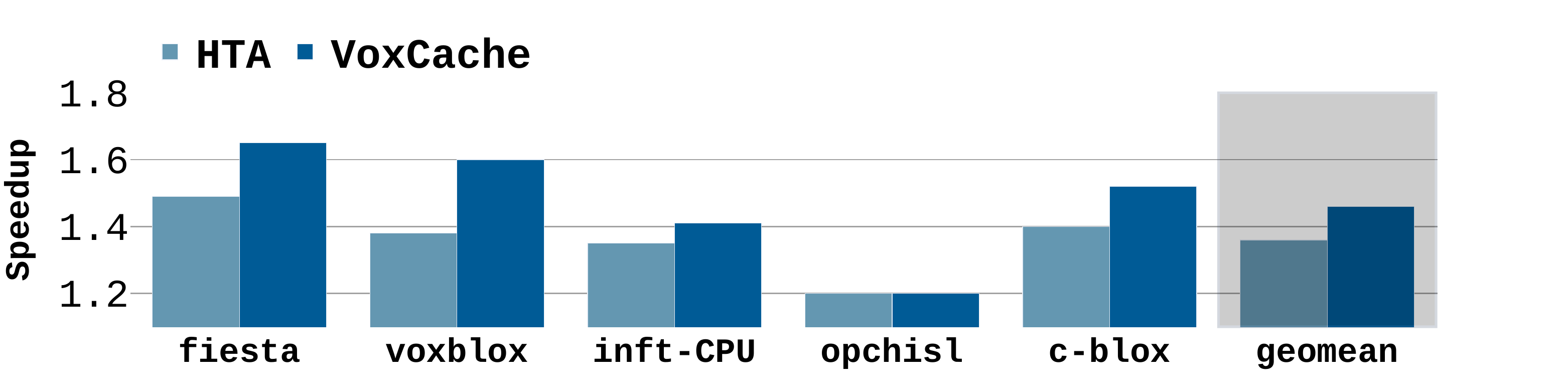}
    \caption{Comparing overall speedup between \Xlabel{}-cached key of CPU workloads vs HTA}
    \label{fig: hta comparison}
    \vspace{-10pt}
\end{figure}

Hash table accelerator~\cite{hta} (HTA) enables fast access to hash table containers by proposing a special hash table storage format in memory. HTA packs hash table keys and values into cache-line sized elements which are stored as an array in memory. A key-value pair's index in the array is derived from the result of a hash function applied on its key. Once a requested cache line is loaded into the cache, HTA is able to achieve fast access with hardware support for key lookup in CPU caches. HTA which uses the same replacement policies for all data used by the application. In contrast to this, our method reserves certain portion of the cache dedicated to key-voxel block pointers, leading to decoupling of map data and other data which our application uses during cache replacement. We compare our results for single threaded CPU workloads with our re-implementation of Flat-HTA, as described in \cite{hta}. We observe that on average, \Xlabel{} has an 8\% higher speedup. The comparison between the speedups achieved between our method and with using HTA is shown in Figure \ref{fig: hta comparison}.

\begin{figure}[htb!]
    \centering
    \includegraphics[width=1.1\linewidth,trim={0 0.2cm 0 0.6cm},clip]{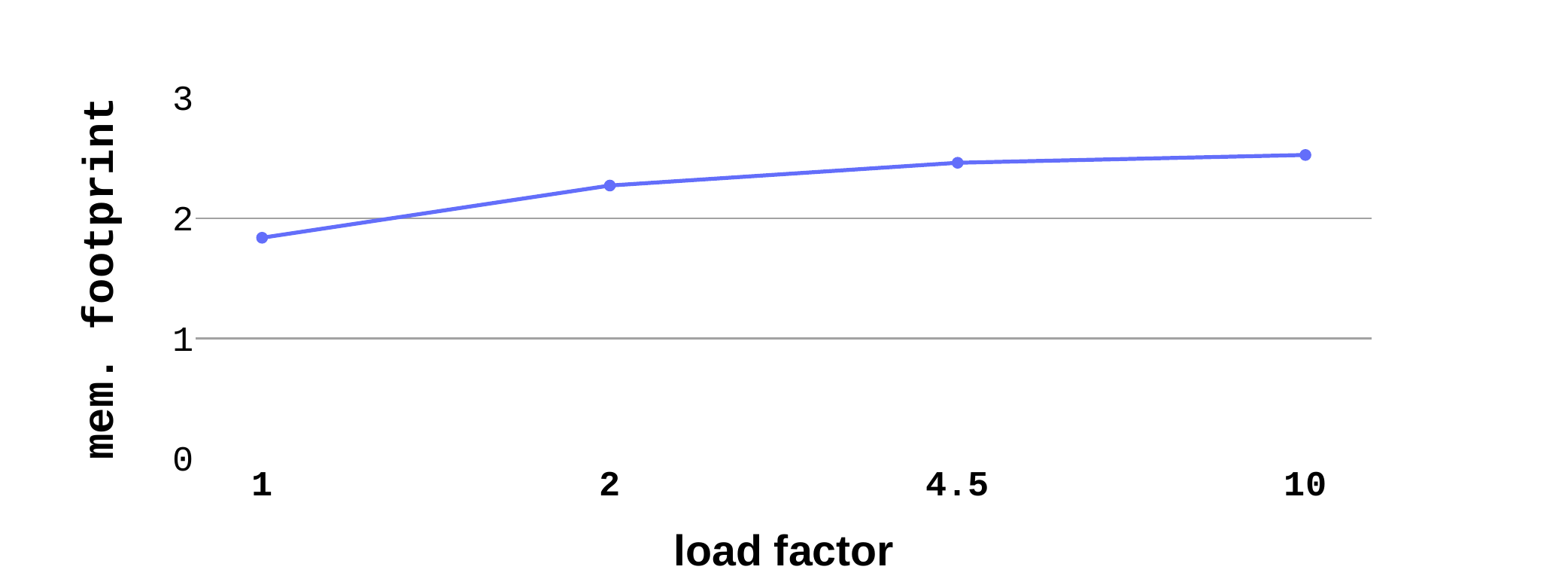}
    \caption{Comparing allocated memory between the baseline vs HTA in CPU workloads for different load factors}
    \label{fig: hta comparison memory}
    \vspace{-10pt}
\end{figure}

We compare the heap allocated memory footprint of HTA and C++ standard library's \texttt{std::unordered\_map} at different load factors, shown in Figure \ref{fig: hta comparison memory}. We observe that HTA would require 2X more memory when compared to the standard library implementation. This approach is thus impractical for large maps in resource-constrained edge devices, e.g., autonomous robots, drones or mobile devices. Our implementation is independent of underlying data structure to store voxel blocks, and enables fast accesses to reusable elements of the hash table regardless of the load factor.

\subsection{\draftmod{\Xlabel{} with Octree Data Structures}}
\draftmodsec{\Xlabel{} caches a pointer to a $n\times n\times n$ block of voxels, and is independent of the data structure used for indexing the pointer to the blocks of voxels. Thus, we can use \Xlabel{} to optimize access latencies in any key-indexed data structure in applications that have a high degree of reuse of keys. In particular, the Supereight~\cite{efficient_largescale_reconstruction} project provides an efficient octree implementation that stores blocks of $8\times8\times8$ voxels as leaves in the octree instead of individual voxels as leaf nodes as an octree. We use \Xlabel{} to further optimize the access times to these blocks of voxels to improve mapping time in Supereight.}

\begin{figure}[htb!]
    \centering
    \includegraphics[width=.85\linewidth,trim={0 0.3cm 0 0.5cm},clip]{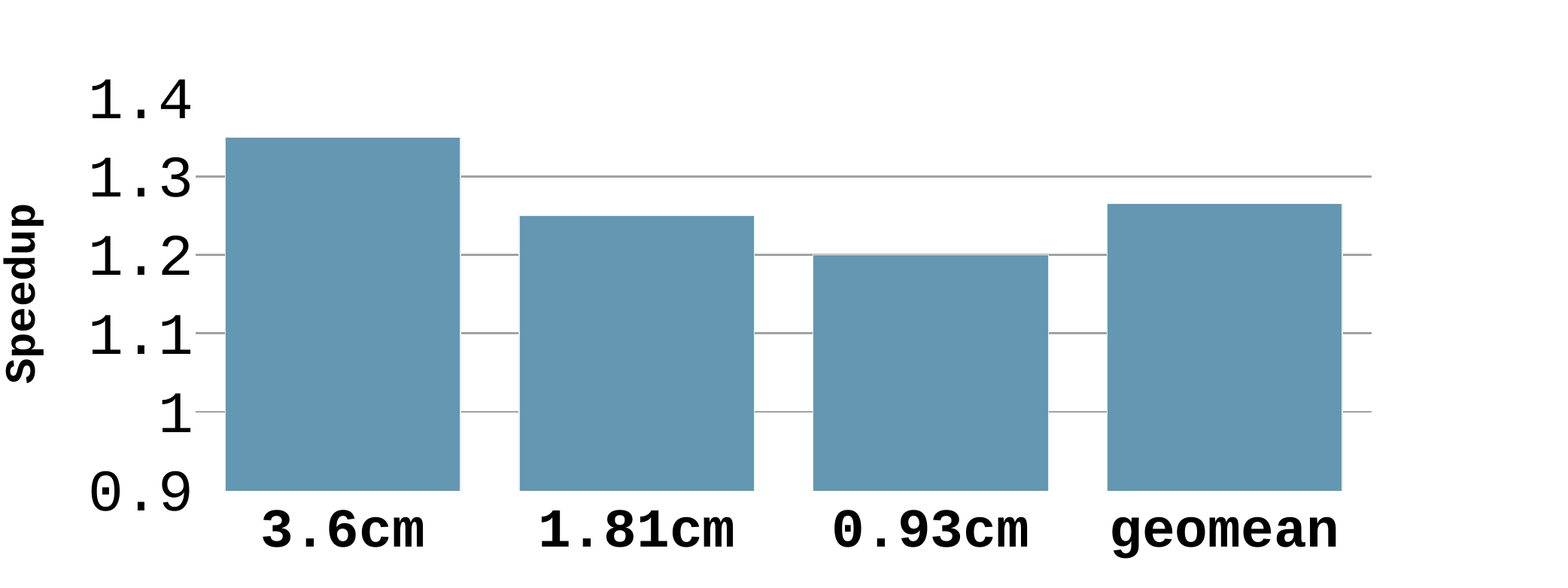}
    \caption{\draftmod{Speedup on mapping update of Supereight with \Xlabel{}}}
    \label{fig:supereight_speedup}
    \vspace{-10pt}
\end{figure}

Figure~\ref{fig:supereight_speedup} shows the speedup obtained on incorporating \Xlabel{} with supereight. We observe that \Xlabel{} enables faster scene integration (of up to $1.31\times$) compared to the baseline implementation. We also note faster raycasting speedups of up to $1.41\times$ on ray casting for mesh generation.

\subsection{\draftmod{Overall Application Performance}}
\Xlabel{} enables efficient computation of high resolution(<5cm) signed distance field on voxels in large environments on the fly. We have evaluated the impact \Xlabel{} for the mapping update step alone in our results. To assess the impact of \Xlabel{} on the overall application, we present the average time taken by InfiniTAM to do iterative-closest-point (localization) and depth integration (mapping) per depth frame input. Table~\ref{table: 5cm impact} and ~\ref{table: 2cm impact} show the results at resolutions 5cm and 2cm respectively. We observe that at a higher resolution, the time contribution of the mapping step increases significantly leading to overall application speedup ($1.24\times$ and $1.31\times$ on CPU and GPU respectively).
\begin{table}[h!]
  \centering
  \caption{\draftmod{Overall application impact of \Xlabel{} at voxel size 2cm}}
  \vspace{-10pt}
  \label{table: 2cm impact}
  \begin{tabular}{|l|l|l|}
    \hline
    \textbf{\Xlabel{}} & \textbf{inft-CPU} & \textbf{inft-GPU}  \\
    \hline
    \hline
     Disabled  & 186 ms & 2.69 ms \\
    \hline
     Enabled  &  150 ms & 2.05 ms  \\
    \hline
  \end{tabular}
    \vspace{-10pt}

\end{table}

\begin{table}[h!]
  \centering
  \caption{\draftmod{Overall application impact of \Xlabel{} at voxel size 5cm}}
  \vspace{-10pt}
  \label{table: 5cm impact}
  \begin{tabular}{|l|l|l|}
    \hline
    \textbf{\Xlabel{}} & \textbf{inft-CPU} & \textbf{inft-GPU}  \\
    \hline
    \hline
     Disabled  & 138 ms & 2.11 ms \\
    \hline
     Enabled  &  127 ms & 2.0 ms  \\
    \hline
  \end{tabular}
  \vspace{-10pt}

\end{table}

\subsection{\Xlabel{} Area Overhead}
The primary area overhead incurred by \Xlabel{} comes \draftmod{from} the additional 3 bits per line on CPUs for the LRU bits and cache state's mode register introduced for each line of the cache. We evaluate the area overhead of \Xlabel{} using CACTI-6.5~\cite{cacti}. In a \draftmod{$32nm$ node} Intel Xeon CPU core, with a 32 KiB L1D, 256 KiB L2 and 1 MiB L3 Cache capacity, on having only the L1D and L2 buffers be able to reserve cache lines, \Xlabel{} incurs an area overhead of at most 0.098\%.

\section{Related Work}
To our knowledge, this is the first work that a) identifies map updates as a significant performance bottleneck and b) proposes a hardware-software mechanism to accelerate large-scale high resolution mapping in 3D reconstruction, SLAM, and robotics applications. Prior work in this area falls into three categories: 1) Faster voxel access methods with special special data structures; 2) Using compression data structures; and 3) Using a faster hash table container. In this section, we discuss each of these types of prior works, and discuss general acceleration techniques on mapping frameworks.

\textbf{Faster voxel access methods.} Fast accesses to voxel data could be achieved using special data structures, like b-trees~\cite{vdb,gvdb} that allow locality aware placement of voxels of close-by regions in memory, resulting in faster map construction~\cite{vdbfusion, vdbedt}. This approach is geared towards storage of large maps to have fewer data movements to and from permanent storage. However, the access times are only comparable to that of voxel hashing~\cite{vdbedt}.

\textbf{Compressing voxel data structures.} Several prior works aim to reduce memory footprint while enabling fast access latencies. Octree based approaches~\cite{octomap,octtree_fusion_scalable,octtree_fusion,efficient_largescale_reconstruction} allow memory efficient representation by compressing sparse regions of space. However, a tree lookup to access voxel data adds significant overhead leading to larger access times compared to voxel hashing~\cite{voxel_hashing}.

\textbf{Fast voxel hash index.} Faster accesses to hash tables for voxel hashing can be achieved by decreasing the load factor of the hash table. This leads to fewer comparisons and less branching required for each hash table lookup. However, this increases the memory footprint making it infeasible to store larger scenes. For example, Google's dense\_hash\_map provides faster accesses compared to \texttt{C++ std::unordered\_map} but requires on average $4X$ the memory required. Hash Table Accelerator (HTA)~\cite{hta} proposes a special hash table storage format in memory enabling fast accesses to hash table containers. HTA packs key-value pairs into cache-line sized elements stored as an array in memory, and indexed on the result of a hash function applied on its key. With hardware support to derive the HTA array index from the key, HTA is able to achieve faster hash table accesses in CPUs. However, HTA will only apply at low load factors, as each bucket can only accommodate keys and values into cacheline sized memory in DRAM, leading to higher memory use by the container. We compare quantitatively the results of \Xlabel{} with our implementation of HTA in Section~\ref{sec:comparison_hta}. The above works trade off scalability of map representation for speed of accesses. Furthermore, HTA is not applicable to frameworks that use other data structures (e.g., octrees) to save voxel data. 

\textbf{Accelerating Mapping and Reconstruction.} 3D mapping, SLAM and 3D reconstruction tasks consist of a number of common sub-problems like localization, scene integration, place recognition, loop closing, and meshing and rendering. \draftmod{Benchmarking tools to estimate contribution of individual sub-problems to the overall runtime for various SLAM projects have been developed~\cite{slambench, illixr}}. To speed up mapping in particular, a number of approaches have been proposed. One class of approaches use properties of TSDF updates to reduce the number of voxel accesses and updates that need to be done. For example, Flash fusion~\cite{flashfusion} stores a lookup table to the addresses of neighboring blocks of voxels to enable faster accesses when performing mesh extraction and TSDF fusion. It also proposes a sampling technique to select only valid voxels which need to be accessed and updated (by discarding accesses to the blocks far away from the truncation distance). However, this would not speed up TSDF fusion in cases where the truncation distance is higher, in which case there are more accesses to the blocks in the camera view frustum. Hierarchical voxel block hashing~\cite{voxhash_hierarchical}, introduces a measure to have an adaptive voxel grid resolution in different parts in the environment, depending on detail. While this allows for a more memory efficient representation, it incurs a higher overhead leading to a lower map update rate. Compared to these approaches, \Xlabel{} is largely orthogonal and can be used in conjunction with these methods, and unlike these approaches \Xlabel{} is more generally applicable as it does not make any assumptions about the environment and is effective at any resolution.

\section{Conclusion}
In this paper we present VoxCache, a hardware-software mechanism to enable fast large scale mapping at high resolutions for SLAM and reconstruction applications. The key idea of our approach is to leverage temporal locality of in accesses to voxel blocks and cache voxel block pointers indexed by their coordinate in space in the cache. We provide hardware support to enable fast lookup of recently accessed voxel blocks from the cache given its key. With our approach, we demonstrate speedups of up to $1.45X$ on CPU and $1.78X$ on GPU on average for mapping frameworks at a high resolution.

\bibliographystyle{ACM-Reference-Format}
\bibliography{refs}

\end{document}